\documentclass{ws-ijmpe}
\usepackage[super,compress]{cite}
\usepackage{hyperref}
 \usepackage{slashed}
\newcommand{\beq}{\begin{equation}}
\newcommand{\beql}[1]{\begin{equation}\label{#1}}
\newcommand{\eeq}{\end{equation}}
\def\bal#1\gal{\begin{align}#1\end{align}}
\newcommand{\ball}[1]{\bal\label{#1}}
\newcommand{\eq}[1]{(\ref{#1})}
\newcommand{\fig}[1]{Fig.~\ref{#1}}
\renewcommand{\sec}[1]{Sec.~\ref{#1}}
\usepackage{yfonts}
\usepackage{tensor} 
\DeclareMathOperator{\tr}{\mathrm{tr}}
\DeclareMathOperator{\Tr}{\mathrm{Tr}}
\DeclareMathOperator{\im}{\mathrm{Im}}
\DeclareMathOperator{\real}{\mathrm{Re}}

\renewcommand{\b}[1]{{\bm #1}} 
\newcommand{\unit}[1]{\hat {{\bm #1}}} 

\newcommand{\e}{\varepsilon}
\newcommand{\aver}[1]{\left\langle #1 \right\rangle}

\begin{document}

\markboth{Jeremy Hansen and Kirill Tuchin}{Chiral effects on radiation and energy loss in quark-gluon plasma}

\catchline{}{}{}{}{}




\title{Chiral effects on radiation and energy loss in quark-gluon plasma}

\author{Jeremy Hansen and Kirill Tuchin}

\address{Department of Physics and Astronomy, Iowa State University, Ames, Iowa, 50011, USA\\
tuchink@gmail.com}

\maketitle

\begin{history}
\received{Day Month Year}
\revised{Day Month Year}
\end{history}

\begin{abstract}

The emergent axion excitations in the quark-gluon plasma are generated by the sphaleron transitions. Their interactions with the photon and gluon fields are  governed by the axion electro- and chromodynamics respectively. We discuss the effect of the emergent axion on the photon production and energy loss in the quark-gluon plasma. In particular, we review the Chiral Cherenkov effect, and the impact of the chiral anomaly on bremsstrahlung.

\end{abstract}



\tableofcontents

\section{Introduction}\label{sec:Intro}

The quark-gluon plasma---hot nuclear matter produced in relativistic heavy-ion collisions---is believed to contain the topological $CP$-odd domains created by the random sphaleron-mediated transitions between different QCD vacua. Interaction of the electromagnetic field with these domains can be described by adding to the QED Lagrangian the  axion-photon coupling term  \cite{Wilczek:1987mv,Carroll:1989vb,Sikivie:1984yz,Fujikawa:2004cx} 
\ball{I1}
\mathcal{L}_A =-\frac{c_A}{4}\theta F_{\mu\nu}\tilde F^{\mu\nu}\,,
\gal
where $c_A= N_c\sum_f q_f^2e^2/2\pi^2$ is the QED anomaly coefficient, $\tilde F_{\mu\nu}= \frac{1}{2}\epsilon_{\mu\nu\lambda\rho} F^{\lambda\rho}$ is the dual field tensor, and the field $\theta$ is sourced by the topological charge density
\ball{I2}
q(x)= \frac{g^2}{32\pi^2}G_{\mu\nu}^a\tilde G^{a\mu\nu}(x) 
\gal
which varies in space and time across a $CP$-odd domain. Assuming  that $\theta$ is slowly varying inside a $CP$-odd domain, it is proportional to the  topological number density \cite{Tuchin:2019jxd}. As a result, \eq{I1} cannot be written as the total derivative and removed from the Lagrangian. Instead, it induces novel terms in the modified Maxwell equations, also known as the Maxwell-Chern-Simons equations, which are proportional to the spatial and the temporal derivatives of $\theta$:
\begin{subequations}\label{I4}
\bal
&\partial_\mu F^{\mu\nu}= j^\nu- c_A \tilde F^{\mu\nu}\partial_\mu\theta\,,\label{I5}\\
&\partial_\mu \tilde F^{\mu\nu}= 0\,.\label{I6}
\gal
\end{subequations}
The same equations describe electromagnetic field in the presence of the axion---a hypothetical field introduced to solve the strong $CP$ problem. However, unlike the fundamental axion, the axion in the hot nuclear matter  is emergent. 
Similar emergent interaction arises in condensed matter systems, in particular in the Dirac and Weyl semimetals due to separation of the chiral modes in momentum space\cite{nenno2020axion}.  The phenomenological manifestations of the axion-photon interaction \eq{I1} were reviewed by many authors, see e.g.\ \cite{Sikivie:2020zpn}. The goal of this review is to focus on a few applications relevant to the quark-gluon plasma. 

The time-dependence of $\theta$ arises due to the finite rate of the sphaleron transitions, finite quark masses and the helicity flow between the magnetic field and QGP.  All these effects have very long characteristic time scales compared to the plasma lifetime \cite{Hirono:2015rla,Tuchin:2014iua,Bodeker:1998hm,Arnold:1998cy,Grabowska:2014efa}, which justifies treating the axion field as slowly varying with time. That the spatial gradients of the axion field are small is indicated by the parametrically large sphaleron size as well as the large correlation length associated with the $CP$-odd domains \cite{Zhitnitsky:2013hs,Zhitnitsky:2012ej}. Thus, all existing models assume the spatial homogeneity of the axion field, while taking account of its slow time variation.\footnote{The impact of domain walls on the chiral magnetic effect in hot QCD matter was analysied in \cite{Tuchin:2018rrw}. } In Weyl semimetals the situation is exactly opposite as we briefly discuss in the concluding section of this review.  
 
We can render \eq{I5} and \eq{I6} explicitly in terms of the electric and magnetic fields:
\begin{subequations}\label{I10}
\bal
&\b \nabla \times \b B = \partial_t \b D + \b j +c_A\partial_t\theta \b B+c_A\b \nabla\theta\times \b E  \,,\label{I11}\\
&\b \nabla\cdot \b D= \rho -c_A\b\nabla\theta \cdot \b B\,,\label{I12}\\
&\b \nabla \times \b E =-\partial_t \b B\,,\label{I13}\\
&\b \nabla\cdot \b B=0\,,\label{I14}
\gal 
\end{subequations}
where we introduced the displacement vector $\b D$ to take account of the medium electric response. The magnetic response is neglected. The new terms that appear in the right-hand-side of Amper's law \eq{I11} are the chiral magnetic  $\b j_\text{CM}= c_A\partial_t\theta\b B= \sigma_\chi \b B$ and the anomalous Hall $\b j_\text{AH}= c_A\b \nabla\theta\times \b E$ currents. Throughout this review we neglect the spatial gradients of the axion field and consider only the chiral magnetic current with constant  chiral magnetic conductivity  $\sigma_\chi$ (also denoted $b_0$). It describes the Chiral Magnetic Effect \cite{Kharzeev:2004ey,Kharzeev:2007tn,Fukushima:2008xe,Kharzeev:2009fn,Kharzeev:2007jp} which is the induction--- by the way of the chiral anomaly \cite{Adler:1969gk,Bell:1969ts}---of anomalous electric current flowing in the magnetic field direction.


\section{Electromagnetic field of a point charge in chiral matter} \label{sec:Z}

In this section we study the electromagnetic fields produced by fast charges in the chiral medium. There are two aspects of these fields that are relevant to the relativistic heavy-ion phenomenology: (i) the electric charges of the incident heavy-ions generate the electromagnetic field in the interaction region and (ii) an electric charge moving through the quark-gluon plasma looses a significant part of its energy by exciting the axion modes in the plasma. While the classical approach is a reasonable approximation in (i), it is too crude for computation of the energy loss in (ii). Nevertheless it provides important insights into the dynamics of the process which will serve us well when we perform a fully quantum calculation in the next section.


Consider a point particle of mass $m$ carrying electric charge $q$  moving in the positive $z$ direction with constant velocity $v$. We assume that the particle energy is much larger than the typical ionization energy, so that the medium response can be described using the macroscopic electromagnetic field. The charge and current densities associated with the particle are $(\rho,\b j)= q \delta(z-vt)\delta (\b b)(1,v\unit z)$, where $\b b$ denotes the transverse components of the position vector $\b r$. The solution to \eq{I10} with $\b D_\omega =\epsilon (\omega)\b E_\omega$, where $E_z= \frac{1}{2\pi}\int_{-\infty}^{\infty} E_{z\omega}e^{-i\omega t}d\omega$ etc., was derived in \cite{Tuchin:2020pbg} as a superposition of the helicity states $\lambda=\pm 1$:
\begin{subequations}\label{Zp4}
\bal
  \b B(\b r,t)= \int \frac{d^2k_\bot d\omega}{(2\pi)^3}e^{i\b k\cdot \b r-i\omega t}\sum_\lambda\b\epsilon_{\lambda \b k} \frac{q \unit z \cdot\b \epsilon_{\lambda\b k}^*\lambda k}{k_\bot^2+\omega^2(1/v^2-\epsilon)-\lambda \sigma_\chi k }\,,\label{Zp6}
 \\
  \b E(\b r,t)= \int \frac{d^2k_\bot d\omega}{(2\pi)^3}e^{i\b k\cdot \b r-i\omega t}\bigg(  \sum_\lambda\b\epsilon_{\lambda \b k}\frac{iq\omega \unit z \cdot\b \epsilon_{\lambda\b k}^* }{k_\bot^2+\omega^2(1/v^2-\epsilon)-\lambda \sigma_\chi k }\nonumber\\
  + \unit k \frac{q}{ivk\e}\bigg)\,, \label{Zp7}
 \gal
 \end{subequations}
 where $\b k = \b k_\bot + (\omega/v)\unit z$ is the wave vector,   $k= \sqrt{k_\bot^2+\omega^2/v^2}$ its length and  $\b\epsilon_{\lambda \b k}$ are the circular polarization vectors satisfying the conditions  $\b\epsilon_{\lambda \b k}\cdot\b\epsilon_{\mu \b k}^*=\delta_{\lambda\mu}$, $\b\epsilon_{\lambda \b k}\cdot \b k=0$ and the identity $
 i\unit k \times\b \epsilon_{\lambda \b k }= \lambda\b \epsilon_{\lambda \b k }$.
Summations over $\lambda$ are performed using the polarization sums given in Appendix of \cite{Tuchin:2020pbg}.  
To compute the energy loss we need only the frequency components of the fields, see \eq{Zb2}. These can be computed exactly\cite{Hansen:2023wzp}:
\begin{subequations}\label{ZA10}
\bal
B_{\phi\omega}(\b r)=& \frac{q}{2\pi}\frac{e^{i\omega z/v}}{k_1^2-k_2^2}\sum_{\nu = 1}^2(-1)^{\nu+1} k_\nu(k_\nu ^2-s^2)K_1(bk_\nu)\,,\label{Za11}\\
B_{b\omega}(\b r)=& \sigma_\chi\frac{q}{2\pi}\frac{i\omega }{v}\frac{e^{i\omega z/v}}{k_1^2-k_2^2}\sum_{\nu = 1}^2(-1)^\nu k_\nu K_1(bk_\nu)\,,\label{Za12}\\
B_{z\omega}(\b r)=& \sigma_\chi\frac{q}{2\pi}\frac{e^{i\omega z/v}}{k_1^2-k_2^2}\sum_{\nu = 1}^2 (-1)^{\nu+1} k^2_\nu K_0(bk_\nu )\,,\label{Za13}\\
E_{z\omega}(\b r)=& \frac{q}{2\pi}\frac{i\omega }{v^2\epsilon}\frac{e^{i\omega z/v}}{k_1^2-k_2^2}\sum_{\nu = 1}^2(-1)^{\nu+1}  \left[(v^2\epsilon-1)(k_\nu^2-s^2)-\sigma_\chi^2\right]K_0(bk_\nu)\,,\label{Za14}\\
E_{b\omega}(\b r)=& \frac{q}{2\pi}\frac{1}{v\epsilon}\frac{e^{i\omega z/v}}{k_1^2-k_2^2}\sum_{\nu =1}^2(-1)^{\nu+1}  k_\nu\left(k_\nu^2-s^2-\sigma_\chi^2\right)K_1(bk_\nu)\,,\label{Za15}\\
E_{\phi\omega}(\b r)=&vB_{b\omega}(\b r)\,,\label{Za16}
\gal
\end{subequations}
 where 
 \ball{Za20}
 k_\nu^2= s^2-\frac{\sigma_\chi^2}{2}+(-1)^\nu \sigma_\chi \sqrt{\omega^2\epsilon+\frac{\sigma^2_\chi}{4}}
 \gal
with $\nu=1,2$ and
\ball{Za21}
s^2=\omega^2\left(\frac{1}{v^2}-\epsilon(\omega)\right)\,.
\gal
Without loss of generality we assume that $\sigma_\chi>0$ which implies that  $k_2^2>k_1^2$. 
The plasma permittivity is well described by
\ball{Za23}
\epsilon =1-\frac{\omega_p^2}{\omega^2+i\omega\Gamma}\,,
\gal
 where $\omega_p$ is the plasma frequency and the damping constant $\Gamma$ is related to the electrical conductivity. 
 

The Fourier integrals over $\omega$ can be computed analytically in the low frequency limit $\epsilon\approx 1+i\sigma/\omega$, where $\sigma$ is the electrical conductivity\cite{Tuchin:2014iua,Li:2016tel,Tuchin:2020pbg}. If one is only interested to compute the magnetic field produced by the relativistic heavy-ions, then one can use the so-called ``diffusion" approximation, which applies to the light-cone times $x_-=t-z/v$ such as
\ball{Zg13}
 x_-\gg \frac{1}{\sigma v^2\gamma^2}\,,\quad x_-\gg \frac{b}{v\gamma}\,,
\gal
where  $\gamma = E_\text{CM}/m_N$, with $E_\text{CM}$ being the center-of-mass energy per nucleon pair and $m_N$ is nucleon mass. For a heavy-ion collision at $\gamma=100$ we estimate $1/\sigma v^2 \gamma^2\sim 3\cdot 10^{-3}$~fm, where we used $\sigma = 5.8$~MeV \cite{Ding:2010ga,Aarts:2007wj,Amato:2013oja,Cassing:2013iz}. For $b\sim 10$~fm, $b/\gamma \sim 0.1$~fm. Thus, \eq{Zg13} holds for the entire lifetime of the hot nuclear medium, which emerges not earlier than $1/Q_s\approx 0.2$~fm after the collision. The magnetic field then reads\cite{Tuchin:2014iua}
\bal
\b B&= -ie\int\frac{d\omega}{2\pi}\int \frac{d^2k_\bot}{(2\pi)^2}
\frac{k_\bot \unit \psi (i\omega \sigma - k_\bot^2)+i\sigma_\chi (\b k_\bot \omega-k_\bot^2\unit z)}{(\sigma^2+\sigma_\chi^2)(\omega-\omega_1)(\omega-\omega_2)}e^{-i\omega x_-+i \b k_\bot \cdot \b b}\label{Zg18}\\
&=\int\frac{d^2k_\bot}{(2\pi)^2} e^{i \b k_\bot \cdot \b b}\int_{-\infty}^{+\infty} \frac{d\omega}{2\pi} \frac{\b f(\omega)}{(\omega-\omega_1)(\omega-\omega_2)}e^{-i\omega x_-} \label{Zg19}\,,
\gal
where we denoted
\ball{Zg20}
\b f(\omega)= -\frac{ie}{\sigma^2+\sigma_\chi^2}
\left[k_\bot \unit \psi (i\omega \sigma - k_\bot^2)+i\sigma_\chi (\b k_\bot \omega-k_\bot^2\unit z)\right]\,
\gal
and 
\ball{Zg17}
\omega_{1,2}= \frac{-i\sigma k_\bot^2\pm k_\bot \sigma_\chi \sqrt{k_\bot^2-\sigma^2-\sigma_\chi^2}}{\sigma^2+\sigma_\chi^2}\,,
\gal 
where $\psi$ is the azimuthal angle in the transverse plane. 
 Closing the  integration contour in  \eq{Zg19}  by an infinite semi-circle in the lower half-plane we find
 at $x_->0$
\bal
\b B =& \int\frac{d^2k_\bot}{(2\pi)^2} e^{i \b k_\bot \cdot \b b}\frac{i}{\omega_2-\omega_1}\left[ e^{-i\omega_1 x_-}\b f(\omega_1)\theta(k_\bot-\sigma_\chi)- e^{-i\omega_2 x_-}\b f(\omega_2)\right]\theta(x_-)\,. \label{Zg21}
\gal
 The value of $\sigma_\chi$ probably does not exceed a few MeV at best, while typical $k_\bot$ is in the range $20-200$~MeV corresponding to $b$'s in the range $1-10$~fm. Therefore, only the case   $k_\bot^2\gg \sigma^2+\sigma_\chi^2$ has a practical significance. This allows us to approximate the poles of \eq{Zg17} as follows
\ball{Zg23}
\omega_{1,2}\approx  \frac{k_\bot^2(-i\sigma\pm \sigma_\chi)}{\sigma^2+\sigma_\chi^2}=\frac{k_\bot^2}{i\sigma\pm \sigma_\chi}\,.
\gal
The magnetic field at  $x_->0$ becomes
\ball{Zg25}
\b B\approx   \int\frac{d^2k_\bot}{(2\pi)^2} e^{i \b k_\bot \cdot \b b}\frac{i}{\omega_2-\omega_1}\left[ e^{-i\omega_1 x_-}\b f(\omega_1)- e^{-i\omega_2 x_-}\b f(\omega_2)\right]\,.
\gal
Performing the remaining integrals yileds
\begin{subequations}
\bal
B_\phi=\frac{eb}{8\pi x_-^2}e^{-\frac{b^2\sigma}{4x_-}}\left[
\sigma \cos\left( \frac{b^2\sigma_\chi}{4x_-}\right)+\sigma_\chi\sin \left(\frac{b^2\sigma_\chi}{4x_-}\right)\right]\,. \\
B_r=\frac{eb}{8\pi x_-^2}e^{-\frac{b^2\sigma}{4x_-}}\left[
\sigma \sin\left( \frac{b^2\sigma_\chi}{4x_-}\right)-\sigma_\chi\cos \left(\frac{b^2\sigma_\chi}{4x_-}\right)\right]\,. 
\\
B_z= -\frac{e}{4\pi x_-}e^{-\frac{b^2\sigma}{4x_-}}\left[
\sigma \sin\left( \frac{b^2\sigma_\chi}{4x_-}\right)-\sigma_\chi\cos \left(\frac{b^2\sigma_\chi}{4x_-}\right)\right]\,.
\gal
\end{subequations}
It is seen that at finite chiral magnetic conductivity $\sigma_\chi$ the magnetic field acquires  $B_r$ and $B_z$ components  and harmonically oscillates as with $b^2/x_-$.


\section{Chiral Cherenkov radiation and energy loss}\label{sec:X}


\subsection{Collisional energy loss in Fermi's model}\label{Zsec:b}

A fast charged particle moving through a medium experiences energy loss due to its interaction with medium particles.  The classical calculation  of the collisional energy loss in the non-chiral medium was first preformed by Fermi \cite{Fermi:1940zz}.  We generalized the Fermi model  to the chiral medium with anomalous response to the magnetic field in \cite{Hansen:2020irw}, utilizing the expressions for the electromagnetic field of the previous section. In this section we review the main results of this classical approach. 

The energy loss rate can be computed as the flux of the Poynting vector out of a cylinder of radius $a$ coaxial with the particle path. For a particle moving with velocity $v$ along the $z$-axis the total loss per unit length reads
\ball{Zb2} 
-\frac{dE}{dz}= 2a\real\int_{0}^{\infty}(E_{\phi\omega} B^*_{z\omega}- E_{z\omega}B^*_{\phi\omega})d\omega\,.
\gal
The real part of the integral arises from the pole in $1/\epsilon$ at $\omega=\omega_p$ as well as from the frequency intervals satisfying $k_1^2<k_2^2<0$ or $k_1^2<0<k_2^2$. Section~III of \cite{Hansen:2020irw} lists the 
corresponding expressions for the integrand of \eq{Zb2}.

\subsubsection{Ultrarelativistic limit}\label{Zsec:b2}

In the limit $ak_\nu\ll 1$ the energy loss reads
\ball{Zb20}
-\frac{dE}{dz}= \frac{q^2}{4\pi}\frac{\omega_p^2}{v^2}\ln \frac{1.12 v}{a\omega_p}- \frac{q^2}{4\pi}\int_{k_1^2<0<k_2^2}\frac{\omega}{v^2\epsilon}\frac{(s^2-k_1^2)(v^2\epsilon-1)+\sigma_\chi^2}{k_1^2-k_2^2}d\omega\,.
\gal
The integration limit  simplifies in the ultrarelativistic limit $v\to 1$: $\omega_p^2/\sigma_\chi<\omega< \gamma^2\sigma_\chi$, where $\gamma=(1-v^2)^{-1/2}$. Expanding the integrand at large frequencies, assuming $\omega\gg \sigma_\chi$, yields
\ball{Zb22}
- \frac{q^2}{4\pi}\frac{1}{2\sigma_\chi}\int^{\gamma^2\sigma_\chi}
\left(- \frac{\sigma_\chi \omega}{\gamma^2}+\sigma_\chi^2\right) d\omega\,,
\gal
where the precise value of the lower limit is irrelevant as long as $\gamma\gg 1$. Integrating over $\omega$ one obtains
\ball{Zb25}
-\frac{dE}{dz}= \frac{q^2}{4\pi v^2}\left( \omega_p^2\ln \frac{v}{a\omega_p}+ \frac{1}{4}\gamma^2\sigma_\chi^2\right)\,.
\gal
The first term on the right-hand-side was obtained by Fermi. We observe that the energy loss due to the anomaly, represented by the second term in \eq{Zb25}, increases as $E^2$ and thus dominates at high energies. In the next subsection we show that quantum effects reduce this dependence to only first power of energy. Nevertheless it still increases faster than the non-anomalous term.

\subsubsection{Non-chiral medium}\label{Zsec:b3}

In the non-chiral limit $\sigma_\chi\to 0$; at finite $a$  the energy loss is given by
\ball{Zb30}
-\frac{dE}{dz}= \frac{q^2}{4\pi}\frac{\omega_p^2}{v^2}K_0(a\omega_p/v)(a\omega_p/v) K_1(a\omega_p/v)
+\frac{q^2}{4\pi v^2}\int_{s^2<0} \omega \left( v^2-\frac{1}{\epsilon}\right)d\omega \,.
\gal
The second term vanishes in completely ionized plasma since $\epsilon<1$ implies that $s^2$ is always positive, see  \eq{Za21} and \eq{Za23}. However, if  medium contains bound states, then the second term contributes when the velocity of the particle is larger than the phase velocity of light in the medium. A single bound state of frequency $\omega_0$ contributes to the permittivity as 
\ball{Zb31}
\epsilon(\omega) = 1-\frac{\omega_p^2}{\omega^2-\omega_0^2+i\omega \Gamma}\,.
\gal
In this case \eq{Zb30} is generalized as
\ball{Zb33}
-\frac{dE}{dz}=& \frac{q^2}{4\pi}\frac{\omega_p^2}{v^2}K_0\left(a\sqrt{\omega_p^2+\omega_0^2}/v\right)\left(a\sqrt{\omega_p^2+\omega_0^2}/v\right) K_1\left(a\sqrt{\omega_p^2+\omega_0^2}/v\right)\nonumber\\
&+\frac{q^2}{4\pi v^2}\int_{s^2<0} \omega \left( v^2-\frac{1}{\epsilon}\right)d\omega \,.
\gal
Neglecting $\Gamma$, the integration region $s^2<0$ is equivalent to  $(1-\epsilon(0)v^2)/(1-v^2)<\omega^2/\omega_0^2<1$ if $v<1/\sqrt{\epsilon(0)}$ and to $\omega<\omega_0$ if $v>1/\sqrt{\epsilon(0)}$. Integration over $\omega$ in the second term yields the well-known Fermi's result \cite{Fermi:1940zz}.

%
\subsubsection{Cherenkov radiation}\label{Zsec:k}

Some of the collisional energy loss emerges in the form of the Cherenkov radiation. In the non-chiral medium it is included in the second term in \eq{Zb30} and is small compared to the large first term that describes medium polarization. 

In the chiral medium the Cherenkov radiation emerges even when $\epsilon=1$, which is known as the chiral (or, in a different context, vacuum) Cherenkov radiation \cite{Tuchin:2018sqe,Huang:2018hgk,Lehnert:2004hq,Lehnert:2004be}.\footnote{It was proposed to be a test of the Lorentz symmetry violation in \cite{Carroll:1989vb,Lehnert:2004hq,Lehnert:2004be,Klinkhamer:2004hg,Mattingly:2005re,Kostelecky:2002ue,Jacobson:2005bg,Altschul:2006zz,Altschul:2007kr,Nascimento:2007rb}.} It is generated by the anomalous electromagnetic current in the presence of the moving charged particle. To compute  the rate of the Chiral Cherenkov radiation emitted in a unit interval of frequencies by an ultrarelativistic particle at $\epsilon=1$ we set $a\to \infty$ in \eq{Zb2} and obtain\cite{Hansen:2020irw}
\ball{Zk2}
\frac{dW}{d\omega} =
\frac{q^2}{4\pi} \left\{ \frac{1}{2}\left( 1-\frac{1}{v^2}\right)+\frac{\sigma_\chi}{2\omega}+\frac{(1+v^2)\sigma_\chi^2}{8v^2\omega^2}+\ldots\right\}\,, \quad \omega< \sigma_\chi\gamma^2\,.
\gal 
 Eq.~\eq{Zk2} is derived neglecting the fermion recoil which is proportional to $\hbar \omega$. It is a good approximation as long as $\omega_+= \sigma_\chi\gamma^2\ll E$, in other words when $\gamma\ll m/\sigma_\chi$,  where $m$ is the particle mass. 
The total radiated power $P$ is obtained by integrating \eq{Zk2} over $\omega d\omega$. It is dominated by the upper limit so that only the first two terms contribute at large $\gamma$ with the result:
\ball{Zk4}
P= \frac{q^2}{4\pi}\frac{\sigma_\chi^2\gamma^2}{4}\,.
\gal
 We observe that the spectrum \eq{Zk2} is exactly the same as \eq{Zb22} which indicates that \emph{all} energy lost by the ultrarelativistic particle due to the anomalous current is radiated as the Chiral Cherenkov radiation.


\subsection{Chiral Cherenkov radiation in quantum theory}\label{sec:chiral-cherenkov-QFT}

Now that we discussed the electromagnetic radiation and the accompanied energy loss in classical theory with the chiral magnetic current, we turn to  calculating the same effect using the quantum field theory. Quantization of the 
theory \eq{I4} encounters difficulties due to the instability of the modes with wavelengths $\lambda>\sigma_\chi$ 
 \cite{Carroll:1989vb,Adam:2001ma}. The instability triggers a very interesting nonlinear process that transfers the chirality of the medium,  proportional to $\sigma_\chi$, to the magnetic helicity of the electromagnetic field.\cite{Joyce:1997uy,Boyarsky:2011uy,Hirono:2015rla,Xia:2016any,Kaplan:2016drz,Kharzeev:2013ffa,Khaidukov:2013sja,Avdoshkin:2014gpa,Akamatsu:2013pjd,Kirilin:2013fqa,Tuchin:2014iua,Dvornikov:2014uza,Buividovich:2015jfa,Manuel:2015zpa,Sigl:2015xva,Kirilin:2017tdh}. Fortunately, the time scale over which this instability develops is much longer than the time scales associated with the quark-gluon plasma. Therefore, in the present review we simply ignore the theoretical issues with the quantization, and assume that it proceeds along the same lines as in the conventional QED except with different dispersion relation of photon.
 
Working in the radiation gauge $A^0=0$ and $\b \nabla\cdot\b A= 0$ the vector potential obeys the equation
\ball{Xb8}
-\nabla^2\b A= -\partial_t^2\b A+\sigma_\chi (\b \nabla\times \b A)\,.
\gal
It follows that the states with given momentum $\b k$, energy $\omega$ and helicity $\lambda$ (i.e.\ circularly polarized plane waves) obey the dispersion relation 
\ball{Xi1}
k^2=\omega^2-\b k^2=  -\lambda\sigma_\chi|\b k|\,.
\gal 
Thus, depending on the sign of  $\lambda\sigma_\chi $ the photon acquires either real or imaginary mass. As a result, one of the polarization states becomes unstable. Recall that  photon radiation by a charged fermion in vacuum $f(p)\to f(p')+\gamma(k)$ and the cross-channel process of pair production in vacuum $\gamma(k)\to f(p')+\bar f(p)$ are prohibited by the energy and momentum conservation. Indeed  in the rest frame of one of the fermions $k^2= (p\pm p')^2= 2m(m\pm E)$. The right-hand-side never vanishes since $E>m$, whereas in the left-hand-side $k^2=0$. In the chiral matter, the dispersion relation \eq{Xi1} opens the $1\to 2$ scattering channels, viz.\ the pair-production if $k^2>0$ and the photon radiation if $k^2<0$. In a matter with positive $\sigma_\chi$, only the right-polarized photons with $\lambda=+1$  can be radiated, while only the left-polarized photons with $\lambda=-1$ decay and vice-versa in a matter with negative $\sigma_\chi$. 

The goal of this section is to compute the rate of photon emission due to the finite chiral magnetic conductivity $\sigma_\chi$ and the corresponding energy loss. In the next subsection we develop the photon wave function in a chiral medium with anisotropic $\sigma_\chi$ which also allows computing the radiative effects associated with the inhomogeneity of the chiral medium. Our calculation relies  on the following approximations: (i) photons and fermions are ultra-relativistic in the laboratory frame (the one associated with the matter), this means that $E\gg m$, $\omega\gg \sigma_\chi$. Apart from making calculations significantly less bulky, this allows neglecting the effect of the electromagnetic field instability in the infrared region as explained above.  (ii) The matter is assumed to be spatially homogeneous within a $CP$-odd domain. (iii) The photon wavelength is much shorter than the domain size.

\subsubsection{Photon and fermion wave functions}\label{Xsec:a} 

The photon wave function is a solution to the modified wave equation \eq{Xb8}. We are interested in a state with  given momentum $k_z\gg k_\bot$ and energy $\omega\gg \sigma_\chi$. In the zeroth order approximation it is described by the vector potential  
\ball{Xb10}
\b A^{(0)}= \frac{1}{\sqrt{2\omega V}}\,\b  e_\lambda  \,e^{i k_z z-i\omega t }\,,\quad k_z=\omega\,,
\gal
where the polarization vector satisfies  $\b  e_\lambda \cdot \unit z =0$ and $V$ is the normalization volume. It is convenient to use the helicity basis $\b  e_\lambda =(\unit x+i\lambda\unit y)/\sqrt{2}$.  To determine the effect of the chiral anomaly on the photon wave function,  we seek for a solution in the form
\ball{Xb12}
\b A= \frac{1}{\sqrt{2\omega V}}\,(\b  e_\lambda \varphi+\unit z \varphi') \,e^{i \omega z-i\omega t }\,,
\gal
where $\varphi$ and $\varphi'$ are functions of coordinates  slowly varying in the longitudinal ($z$) direction, viz.\ $|\partial_z\varphi/\varphi|\ll \omega $ and  $|\partial_z\varphi'/\varphi'|\ll \omega $. The two unknown functions $\varphi$ and $\varphi'$ are required in order to account for the change of the direction of the  photon polarization. The gauge condition yields a constraint
\ball{Xb13}
(\b e_\lambda \cdot \b \nabla_\bot) \varphi+\partial_z\varphi'+i\omega \varphi'\approx (\b e_\lambda \cdot \b \nabla_\bot) \varphi+i\omega \varphi'=0\,.
\gal
Substituting \eq{Xb12} into \eq{Xb8} one obtains
\ball{Xb14}
\b  e_\lambda \left(-2i\omega \partial_z\varphi-\nabla_\bot^2\varphi\right)+\unit z \left(-2i\omega \partial_z\varphi'-\nabla_\bot^2\varphi'\right)\nonumber\\
= \sigma_\chi\left(\omega \lambda \b e_\lambda \varphi - \b e_\lambda \times \b\nabla \varphi-\unit z\times \b\nabla_\bot \varphi'\right)\,.
\gal
Taking the scalar product of this equation with $\b  e_\lambda ^*$ and  using $\b e_\pm^*\cdot \b e_\mp=0$ and $\b e_\pm^*\cdot \b e_\pm=1$  produces
\ball{Xb16}
-2i\omega \partial_z\varphi-\nabla_\bot^2\varphi= \sigma_\chi (\omega\lambda \varphi-i\lambda \partial_z\varphi)+i\lambda\b e_\lambda ^*\cdot \b \nabla_\bot\varphi'\,,
\gal
where we used the identity $\b e_\lambda\times \unit z= i\lambda \b e_\lambda$.
In view of \eq{Xb13} one can drop the small term proportional to $\varphi'$. Neglecting also $\partial_z\varphi$ in parentheses furnishes the equation for $\varphi$:
\ball{Xb18}
-2i\omega \partial_z\varphi-\nabla_\bot^2\varphi= \sigma_\chi \omega\lambda \varphi\,.
\gal
Taking the scalar product of \eq{Xb14} with $\unit z$ yields
\ball{Xb20}
-2i\omega \partial_z\varphi'-\nabla_\bot^2\varphi'= \sigma_\chi i\lambda(\b e_\lambda \cdot \b \nabla)\varphi\,.
\gal
One can eliminate in \eq{Xb20} the term proportional to $\varphi$ using the gauge condition \eq{Xb13}. This furnishes the equation for $\varphi'$, which is precisely the same as equation \eq{Xb18} obeyed by $\varphi$. 

A solution to \eq{Xb18} can be written as
\ball{Xb22}
\varphi= e^{i\b k_\bot \cdot \b x_\bot}\exp\left\{ -i \frac{1}{2\omega} \int_0^z\left[k_\bot^2-\sigma_\chi(z')\omega \lambda\right]dz'\right\}\,.
\gal
It follows from \eq{Xb13} that 
\ball{Xb24}
\varphi' = -\frac{\b e_\lambda\cdot \b k_\bot}{\omega}\varphi\,.
\gal
Substituting \eq{Xb22} and \eq{Xb24} into \eq{Xb12} yields the photon wave function in the high energy approximation \cite{Tuchin:2018sqe}
\ball{Xb25}
\b A= \frac{1}{\sqrt{2\omega V}}\b  \epsilon_\lambda  \,e^{i \omega z+i\b k_\bot \cdot \b x_\bot-i\omega t }\exp\left\{ -i \frac{1}{2\omega} \int_0^z\left[k_\bot^2-\sigma_\chi(z')\omega \lambda\right]dz'\right\}\,,
\gal
where  the polarization vector 
\ball{Xb26}
\b\epsilon_\lambda = \b e_\lambda - \frac{\b e_\lambda\cdot \b k_\bot}{\omega} \unit z\,.
\gal
Clearly, $\b  \epsilon_\lambda \cdot \b k=0$ up to the terms of order $k_\bot^2/\omega^2$ and $\sigma_\chi/\omega$. 
 If the scattering process happens entirely within a single domain, then the chiral magnetic conductivity is constant. However, if a domain wall is located at, say, $z=0$, then the chiral magnetic conductivity is different at $z<0$ and $z>0$. This is why a possible $z$-dependence of $\sigma_\chi$ is indicated in \eq{Xb25}. Even though the boundary conditions on the  domain wall induce a reflected wave, it can be neglected in the ultra-relativistic approximation \cite{Baier:1998ej,Schildknecht:2005sc}.  
 
To conclude this subsection, we observe that in the ultra-relativistic limit $\omega\gg k_\bot\gg |\sigma_\chi|$ the 
dispersion relation can be cast in the form
\ball{Xa15}
k_z\approx \omega -\frac{1}{2\omega}\left(k_\bot^2-\lambda\sigma_\chi \omega\right)
\gal
which has only real solutions $\omega$. This illustrates the point we made in the beginning of this section, that the instability of the infrared modes does not influence the short-distance processes.


The free fermion wave function $\psi$ at high energy $E\gg p_\bot,m$ can be obtained using the same procedure using the high energy approximation to the Dirac equation: 
\ball{Xb35}
\psi = \frac{1}{\sqrt{2E V}}u(p)  e^{iE( z- t)}\exp\left\{ i\b p_\bot\cdot \b x_\bot-iz\frac{\b p_\bot^2+m^2}{2E}\right\}\,.
\gal

\subsubsection{Photon radiation}\label{Xsec:d}

The scattering matrix element for the photon radiation $f(p)\to f(p')+\gamma(k)$ is calculated according to the standard rules:
\bal
S=& -ie q\int \bar \psi \gamma^\mu \psi A_\mu d^4x\label{d3}\\
=&i(2\pi)^3\delta(\omega+E'-E)\delta(\b p_\bot-\b k_\bot-\b p'_\bot)\frac{\mathcal{M}}{\sqrt{8EE' \omega V^3}}\,,
\label{Xd5}
\gal
where $q$ is the fermion electric charge.  The wave functions $\varphi_k$ and $\phi_p$ are given by \eq{Xb22} and \eq{Xb35} respectively with the subscripts indicating the corresponding momenta. The amplitude  $\mathcal{M}$ is given by 
\bal
\mathcal{M}=& -eQ\bar u(p')\gamma^\mu u(p)\epsilon^*_\mu\nonumber\\
&\times
\int_{-\infty}^\infty dz \exp\left\{ i\int_0^z dz' \left[ \frac{{p'}^2_\bot+m^2}{2E'}-\frac{p_\bot^2+m^2}{2E}+\frac{k_\bot^2-\sigma_\chi\omega\lambda}{2\omega}\right]\right\}
\label{Xd7}\\
=& \mathcal{M}_0\int_{-\infty}^\infty dz \exp\left\{ i\int_0^z  \frac{q_\bot^2+ \kappa_\lambda(z')}{2E x(1-x)}dz'\right\} \,,
\label{Xd8}
\gal 
where we introduced notations $\mathcal{M}_0=-eq\bar u(p')\gamma^\mu u(p)\epsilon^*_\mu$, $x=\omega/E$,
\ball{Xd10}
\b q_\bot = x\b p'-(1-x)\b k_\bot\,,
\gal
and 
\ball{Xd12}
\kappa_\lambda(z) = x^2m^2-(1-x)x\lambda \sigma_\chi E\,.
\gal
Computing the amplitude $\mathcal{M}_0$  one obtains
\bal
\mathcal{M}_0&=-eq\bar u_{\sigma'}(p')\gamma\cdot \epsilon^*_{\lambda}(k) u_{\sigma}(p) \label{Xd28}\\
&= -\frac{eq}{\sqrt{2(1-x)}}\left[ xm(\sigma+\lambda)\delta_{\sigma',-\sigma}
-\frac{1}{x}(2-x+x\lambda \sigma)(q_x-i\lambda q_y)\delta_{\sigma',\sigma}
\right]\label{Xd29}\,,
\gal
where $\sigma=\pm 1$ and $\sigma'\pm 1$ are the fermion helicities before and after photon radiation. 

The transition probability can be computed as
\ball{Xd33}
dw= |S|^2\frac{V d^3p'}{(2\pi)^3}\frac{V d^3k}{(2\pi)^3}\,,
\gal
while the number of produced photons $N$  is the cross section per unit area as
\ball{Xd35}
dN= \frac{1}{(2\pi)^3}\frac{1}{8x(1-x)E^2}\frac{1}{2}\sum_{\lambda,\sigma,\sigma'}|\mathcal{M}|^2d^2q_\bot dx\,,
\gal
where the sum runs over the photon and fermion helicities.

\subsubsection{One infinite domain}

Consider first an infinite chiral matter with constant chiral conductivity. Performing  the integral over $z$ in \eq{Xd8} yields
\ball{Xda1}
\mathcal{M}&= 2\pi \mathcal{M}_0\,\delta\left(\frac{q_\bot^2+\kappa_\lambda}{2E x(1-x)}\right)\,.
\gal
Substituting it into  \eq{Xd35} we derive the spectrum of the photon radiation rate $W$
\ball{Xd50}
\frac{dW}{dx}= \frac{1}{16\pi\e}\frac{1}{2}\sum_{\lambda,\sigma,\sigma'}|\mathcal{M}_0|^2\,\theta(-\kappa_\lambda)\,,
\gal
where $\theta$ is the step-function. In view of \eq{Xd12} that the parameter $\kappa_\lambda$ is negative if $\lambda\sigma_\chi>0$ and 
\ball{Xd52} 
x<x_0=\frac{1}{1+m^2/(\lambda\sigma_\chi E)}\,.
\gal 
Assume for definitiveness that $\sigma_\chi>0$. Then only the right-polarized photons with $\lambda>0$ are radiated. Using \eq{Xda1}, \eq{Xd29} in \eq{Xd50} and performing the  summations and the integration yields the radiation rate for two circular polarization states \cite{Tuchin:2018sqe}: 
\bal
\frac{dW_+}{dx}&= \frac{\alpha q^2}{2\e x^2(1-x)}\left\{ -\left(\frac{x^2}{2}-x+1\right)\kappa_++\frac{x^4m^2}{2}\right\}\theta(x_0-x)
\nonumber\\
&= \frac{\alpha q^2}{2\e x }\left\{ \sigma_\chi \e \left(\frac{x^2}{2}-x+1\right)-m^2x\right\}\theta(x_0-x)\,,
\label{Xd55}\\
\frac{dW_-}{dx}&=0\,.\label{Xd56}
\gal
The photon spectrum radiated in a matter with $\sigma_\chi<0$ can be obtained by  replacing $W_\pm\to W_\mp$ and $\sigma_\chi \to -\sigma_\chi$. Note that since the anomaly coefficient $ c_A\sim  \alpha$, the spectrum \eq{Xd55} is of the order  $\alpha^2$. The total energy radiated by a fermion per unit length $dz$ is\cite{Tuchin:2018sqe}
\ball{Xd58}
\frac{dE}{dz} = \int_0^1 \frac{dW_+}{dx} x\e dx= \frac{1}{3}\alpha q^2\sigma_\chi \e\,,
\gal
where the terms of order $m^2/|\sigma_\chi|E$ have been neglected.

\subsubsection{Two semi-infinite domains separated by a domain wall at $z=0$}\label{Xsec:e2}

Suppose now that the chiral matter consist of two  semi-infinite domains separated by a thin flat domain wall at $z=0$.  Performing  the integral over $z$ in \eq{Xd8} yields  
\bal
\mathcal{M}&=  2E x(1-x)\mathcal{M}_0\left\{ \frac{-i}{q_\bot^2+\kappa'_\lambda-i\delta}-\frac{-i}{q_\bot^2+\kappa_\lambda+i\delta}
\right\}\,, \label{Xd26}
\gal
where the values of $\kappa_\lambda$ at $z<0$ and $z>0$ are denoted by $\kappa'_\lambda$ and $\kappa_\lambda$ respectively and infinitesimal $\delta>0$ is inserted to regularize the integrals.  Plugging \eq{Xd26}, \eq{Xd29} into \eq{Xd35} and performing summation over spins yields the radiation spectrum\cite{Tuchin:2018sqe} 
\ball{Xd40}
\frac{dN}{d^2q_\bot dx}= \frac{\alpha q^2}{2\pi^2 x}\left\{ \left(\frac{x^2}{2}-x+1\right)q_\bot^2+\frac{x^4m^2}{2}\right\}
\sum_\lambda \left| \frac{1}{q_\bot^2+\kappa'_\lambda-i\delta}-\frac{1}{q_\bot^2+\kappa_\lambda+i\delta}\right|^2\,.
\gal
The spectrum peaks at $q_\bot^2=-\kappa_\lambda$ and/or  $q_\bot^2=-\kappa'_\lambda$ provided that $\kappa_\lambda<0$ and/or  $\kappa_\lambda'<0$ respectively. These peaks represent the Chiral Cherenkov radiation. Away from the peaks, one can neglect $\delta$ in \eq{Xd40}. The resulting spectrum coincides with the spectrum of the conventional \emph{transition radiation} once $\kappa_\lambda$'s are replaced by $\kappa_\mathrm{tr}=m^2x^2+m_\gamma^2(1-x)>0$, where  $m_\gamma$ is the effective photon mass  \cite{Baier:1998ej,Schildknecht:2005sc}. The conventional transition radiation is essentially a classical phenomenon. Indeed its spectrum is finite in the limit $\hbar \to 0$. In contrast, the chiral transition radiation is a purely quantum effect because $\sigma_\chi$ vanishes in that limit. 

The integral over the momentum $q_\bot$ in \eq{Xd40} is dominated by the poles at $q_\bot^2=-\kappa_\lambda$ and $q_\bot^2=-\kappa'_\lambda$. There are two distinct cases depending on whether $\sigma_\chi$ and $\sigma_\chi'$ have the same or opposite signs. Consider first $\sigma_\chi>0$ and $\sigma_\chi'>0$. In this case the photon spectrum is approximately right-polarized. Keeping only the terms proportional to $1/\delta$ one obtains
\ball{Xdc1}
\frac{dW_{++}}{dx}= \frac{\alpha q^2}{8 x^2(1-x)E}\left[ \left(\frac{x^2}{2}-x+1\right)|\kappa_++\kappa_+'|+\frac{x^4m^2}{2}\right]
\theta(x_0-x)\theta(x_0'-x)\,,
\gal
where the double plus subscript indicates that the helicity is positive in both domains.
The maximum energy fraction taken by the photon $x_0$ is defined in \eq{Xd52}; $x_0'$ is the same as $x_0$ with $\sigma_\chi$ replaced by $\sigma_\chi'$.   Consider now  $\sigma_\chi'>0$ and $\sigma_\chi<0$. The integration gives
\ball{Xdc2}
\frac{dW_{+-}}{dx}=& \frac{\alpha q^2}{8 x^2(1-x)E}\left\{ \left[\left(\frac{x^2}{2}-x+1\right)|\kappa_+'|+\frac{x^4m^2}{4}
\right] \theta(x_0'-x)\right. \nonumber\\
&
\left.
+\left[\left(\frac{x^2}{2}-x+1\right)|\kappa_-|+\frac{x^4m^2}{4}
\right] \theta(x_0-x)\right\}\,.
\gal
Clearly,  photons radiated to the left of the domain wall ($z<0$)  are right-polarized, while those radiated to its right ($z>0$) are left-polarized.  

We mentioned in the beginning of this section that the energy and momentum conservation prohibit the pair production $\gamma(k)\to \bar f (p)+f(p')$ in vacuum. However, in the  chiral matter this channel opens up. In fact it is the cross channel of the photon radiation. The detailed analysis of this process if performed in \cite{Tuchin:2018sqe}.

In summary, the chiral term in the photon dispersion relation generates the spacelike gluon mode  $\omega^2-\b k^2<0$ which opens the possibility for novel $1\to 2$ processes that are otherwise prohibited in QED by energy and momentum conservation. This effect is analogous to the Cherenkov radiation where the spacelike excitations of the electromagnetic field produced by a particle moving at a speed greater than the phase velocity of light, represent the propagating wave solution in dielectric materials. In chiral media, the chiral conductivity effectively contributes to the medium dielectric response making possible excitation of the gauge field wave. This effect is referred to as the Chiral Cherenkov radiation. Additionally, in the medium with varying topological charge density the chiral transition radiation is also possible. The novel chiral radiation processes  can be used to investigate the chiral anomaly in such diverse media as the quark-gluon plasma, Weyl semimetals, and axionic dark matter \cite{Huang:2018hgk}.


\subsection{Color Chiral Cherenkov radiation and energy loss in quark-gluon plasma}\label{sec:CCC}


Thus far we have discussed the chiral radiation phenomena in QED. In particular we focused on the Chiral Cherenkov radiation emitted by a  fast  particle carrying electric charge and moving in a medium with chiral fermions
\cite{Tuchin:2018sqe,Huang:2018hgk,Hansen:2020irw,Tuchin:2018mte}. It is emitted due to the unique dispersion relation of the electromagnetic field and is closely related to the chiral magnetic effect \cite{Kharzeev:2004ey,Kharzeev:2007jp,Kharzeev:2009fn,Kharzeev:2007tn,Fukushima:2008xe}. 
It is clear that the color version of the Chiral Cherenkov effect is also expected to give a significant contribution to the energy loss in strongly interacting media with chiral fermions. In this subsection we review the QCD version of the Chiral Cherenkov effect developed recently in \cite{Hansen:2024rlj}. 

As in \sec{sec:Intro} we describe the  gluon field excitations in the chiral medium supporting the chiral magnetic current by the Lagrangian \cite{Wilczek:1987mv,Carroll:1989vb,Sikivie:1984yz}
\ball{a1}
\mathcal{L}= -\frac{1}{2} \Tr\left(F_{\mu\nu} F^{\mu\nu}\right)-\frac{c'_A}{2} \theta \Tr\left(  F_{\mu\nu}\tilde F^{\mu\nu}\right)  \,,
\gal
where the field tensor is $F_{\mu\nu}= \partial_\mu A_\nu -\partial_\nu A_\mu-ig[A_\mu, A_\nu]$, 
$A_\mu = A_\mu^at^a$, $F_{\mu\nu}= F_{\mu\nu}^at^a$, and $t^a$ are the SU(3) generators. The pseudo-scalar field $\theta$ is sourced by the topological charge and $c'_A$ is the QCD anomaly coefficient. As in QED we adopt a model of the quark-gluon plasma with  spatially homogeneous and slowly time dependent $\theta$: $c'_A\partial_\mu \theta= b_0\delta_{\mu 0}$, where $b_0$ is the color chiral magnetic conductivity assumed to be constant\footnote{In this subsection we use $b_0$ in place of $\sigma_\chi$ to distinguish between the responses to chromo-magnetic  and magnetic fields respectively.}.
As a result the chiral magnetic current $\b j^a= b_0\b B^a$ emerges as a source of the color magnetic field in Amper's law. That $\theta$ is a slow function of time is indicated by the parametrically large 
sphaleron transition time $1/(g^4 T)$ at a given plasma temperature $T$ \cite{Arnold:1996dy,Arnold:1998cy,Bodeker:1998hm}.

The free  gluon excitations are governed by the equation
\ball{a2}
\partial_\nu F^{\mu\nu}+b_\nu \tilde F^{\mu\nu}=0\,,
\gal
along with the Bianchi identity. In the radiation gauge, the corresponding vector potential obeys the equation
\ball{a3}
\partial_t^2\b A-\nabla^2\b A= b_0 (\b \nabla\times \b A)\,,
\gal
which is color version of \eq{Xb8}. The plane wave solutions of \eq{a3} are circularly polarized and have the same dispersion relation as its Abelian counterpart (cp.\ to \eq{Xi1})
\ball{a4}
\omega^2= \b k^2 -\lambda b_0 |\b k|\,.
\gal
This dispersion relation ignores screening effects which is a justifiable assumption only for axion electromagnetic excitations. With the account of the screening, the in-medium gluons have the following dispersion relation 
\ball{t3}
\omega^2= \b k^2 -\lambda b_0 |\b k|+\omega_p^2\,,
\gal
where $\omega_p$ is the plasma frequency. Actually, Eq.~\eq{t3} is the short wavelength limit of the full dispersion relation \cite{Akamatsu:2013pjd,Manuel:2013zaa}, which nevertheless suffices for the present calculation. To simplify the notation we assume that $b_0$ is positive.

\begin{figure}
    \centering  \includegraphics[width=.7\linewidth]{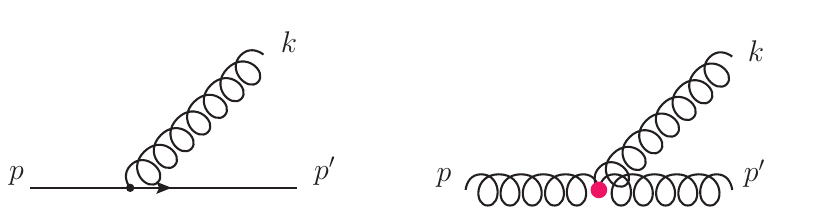}
    \caption{ $1\to 2$ processes contributing to the Color Chiral Cherenkov radiation. The anomalous contributions come about by the way of the gluon dispersion relation \eq{t3} and as an extra term in the  triple-gluon vertex \eq{f5}. The latter fact is indicated by the big red circle.
 }
    \label{fig:channels}
\end{figure}

Unlike QED, where the chrial Cherenkov radiation is described by a single diagram $e\to e\gamma$, where the photon dispersion relation is modified by the anomaly, in QCD there are two possible channels depicted in \fig{fig:channels}. The first of these channels $q\to qg$ is quasi-Abelian.  It was discussed in \sec{Xsec:d} and those results apply in the present case as well up to the overall color factor. The novel channel is $g\to gg$ where all three gluons are excitations of the chiral medium, see \fig{fig:channels}. To derive the corresponding Feynman rule 
we write in the second term in \eq{a1} 
\ball{f1}
\Tr\left( F^{\mu\nu}\tilde F^{\mu\nu}\right)= 2\partial_\mu K^\mu
\gal 
where
\ball{f2}
K^\mu= \epsilon^{\mu\nu\rho\sigma}\Tr\left(A_\nu\partial_\rho A_\sigma-g\frac{2i}{3}A_\nu A_\rho A_\sigma\right)
\gal
and integrate by parts the corresponding term in the action. The result is 
\ball{f4}
S_\theta= c_A\int  \partial_\mu\theta \epsilon^{\mu\nu\rho\sigma}\left(\frac{1}{2}A^a_\nu\partial_\rho A^a_\sigma-g\frac{2i}{3}\frac{1}{4}if^{abc}A^a_\nu A^b_\rho A^c_\sigma\right)d^4x\,.
\gal
The first term in \eq{f4} contributes to the equation of motion \eq{a2}, while the second one to the triple-gluon vertex. The variation of action $S_\theta$ with respect to three gluon fields produces the Feynman rule for the anomalous contribution to the triple-gluon vertex in momentum space\cite{Hansen:2024rlj}:
\ball{f5}
\text{anomalous triple-gluon vertex}=gb_\mu\epsilon^{\mu\nu\rho\sigma}f^{abc}\,,
\gal
where we follow the conventions of  \cite{Peskin:1995ev}. This is the only new Feynman rule due to the anomalous term in the Lagrangian \eq{a1}. The red circle in \fig{fig:channels} includes both the conventional and anomalous contributions.



The radiation rates can be computed as 
\ball{c9}
dW_{a\to bc}= \frac{g^2 C_{a\to bc}}{2(2\pi)^2}\sum_{ss'}\delta(\omega+E'-E)\delta(\b k+\b p'-\b p) \frac{ |\mathcal{M}_{a\to bc}|^2}{8E E' \omega}
  d^3p'\, d^3k\,,
\gal
where the sum runs over the fermion polarization states, $\mathcal{M}_{a\to bc}$ are the amplitudes without the color generators $t^a$ and the structure constants $f^{abc}$, and $C_{a\to bc}$ are the color factors given by
\begin{subequations}
\bal
C_{q\to qg}&= \frac{1}{N_c}\tr(t^a t^a)=\frac{N_c^2-1}{2N_c}=\frac{4}{3}\,,\label{u1}\\
C_{g\to gg}&=\frac{(f^{abc})^2}{N_c^2-1}=N_c=3\,.\label{u2}
\gal
\end{subequations}
The matrix elements read:
\bal
i\mathcal{M}_{q\to qg}&=i\bar u_{\b p' s'}\slashed{e}^*_{\b k \lambda} u_{\b p s}\,,\label{t1}\\
i\mathcal{M}_{g\to gg}&=
(e_{\b p \lambda_0}\cdot e^*_{\b k \lambda})(p+k)\cdot e^*_{\b p' \lambda'}
+(e^*_{\b k \lambda}\cdot e^*_{\b p' \lambda'})(p'-k)\cdot e_{\b p \lambda_0}\nonumber\\
&-
(e_{\b p \lambda_0}\cdot e^*_{\b p' \lambda'})(p+p')\cdot e^*_{\b k \lambda}-ib_\mu\epsilon^{\mu\nu\rho\sigma}e_{\b p \lambda_0,\nu}e^*_{\b k \lambda,\rho}e^*_{\b p' \lambda',\sigma}\,,\label{t2}
\gal
where $e_{\b p\lambda \mu}$'s are the circular polarization vectors. As before we the incident particle moves in the $z$ with ultrarelativistic velocity so that $k_\bot,b_0 \ll E,E',\omega$.

The integral over the $\b p'$ in \eq{c9} is trivial considering the delta-function expressing the momentum conservation. The remaining delta-function in $q\to qg$ rate reads:
\ball{t10}
\delta(E'+\omega-E)&=2x(1-x)E\delta\left[k_\bot^2+(\omega_p^2-\lambda b_0 xE)(1-x)+m^2x^2\right]\,.
\gal
Clearly, the argument of this delta-function can vanish only if $\lambda b_0>0$. We assumed that $b_0>0$. Therefore, only the right-polarized gluons $\lambda>0$ can be radiated by the incident quark. Moreover, the energy conservation, expressed by the argument of the delta-function \eq{t10}, can be satisfied only if 
\ball{t12}
x_-^{q\to qg} <x<x_+^{q\to qg}\,,
\gal
where 
\ball{t13}
x_\pm^{q\to qg}=\frac{\omega_p^2+\lambda b_0 E\pm\sqrt{(\omega_p^2-\lambda b_0 E)^2-4m^2\omega_p^2}}{2(m^2+\lambda b_0 E)}\,.
\gal

The  delta-function in $g\to gg$ rate reads:
\ball{t15}
\delta(E'+\omega-E)&=2x(1-x)E\delta\left[k_\perp^2+\omega_p^2(1-x+x^2)-b_0E(\lambda+\lambda'-\lambda_0)x(1-x)\right]\,.
\gal
In this case the gluon emission is possible only if $\lambda+\lambda'>\lambda_0$. Apparently only the following four channels are allowed: $g_L\to g_Rg_L$, $g_L\to g_Lg_R$, $g_R\to g_Rg_R$, $g_L\to g_Rg_R$. In the first three 
 of these channels $\lambda+\lambda'-\lambda_0=1$, whereas in the last one $\lambda+\lambda'-\lambda_0=3$. Additionally,  
\ball{t18}
x_-^{g\to gg} <x<x_+^{g\to gg}\,,
\gal
where 
\ball{t19}
x_\pm^{g\to gg}=\frac{\omega_p^2+(\lambda+\lambda'-\lambda_0) b_0 E\pm\sqrt{(\omega_p^2-(\lambda+\lambda'-\lambda_0) b_0 E)^2-4\omega_p^4}}{2(\omega_p^2+(\lambda+\lambda'-\lambda_0) b_0 E)}\,.
\gal
After some algebra \eq{c9} yields, aftre summing over the final gluon polarizations,\cite{Hansen:2024rlj} 
\begin{subequations}\label{C2}
\bal
\frac{dW_{q\rightarrow qg}}{d k_\perp^2 d x}=&\frac{\alpha_s g^2}{3x^2 (1-x)E }\left[(1+(1-x)^2) k_\perp^2+m^2x^4\right]\nonumber\\
&\times\delta\left(k_\perp^2+\mu^2 (1-x)+m^2x^2\right) \,,\label{C2a}\\
\frac{dW_{g_R\rightarrow g g}}{d k_\perp^2 d x}=&\frac{3\alpha_s g^2}{2x^2 (1-x)^2E } k_\perp^2\delta\left(k_\perp^2+\omega_p^2x^2+(\omega_p^2-b_0 \omega) (1-x)\right) \,,\label{C2aa}\\
\frac{dW_{g_L\rightarrow g g}}{d k_\perp^2 d x}=&\frac{3\alpha_s g^2 }{2 E}\left[\frac{ (1-x)^2}{x^2 }+\frac{ x^2}{(1-x)^2 }\right] k_\perp^2\nonumber\\
&\times\delta\left(k_\perp^2+\omega_p^2x^2+(\omega_p^2-b_0 \omega) (1-x)\right) \,.\label{C2ab}
\gal
\end{subequations}
One can easily identify in \eq{C2} the contributions to the standard splitting functions $P_{gq}(x)$ and $P_{gg}(x)$ corresponding to gluon emission.
Integrating over the transverse momentum  $k_\bot$ we obtain the spectra of the gluon emission rate: 
\begin{subequations}\label{C11}
\bal
\frac{dW_{q\rightarrow qg}}{d x}=&\frac{\alpha_s g^2}{3 x^2 E}\left\{\left[1+(1-x)^2\right] \left(b_0 xE-\omega_p^2\right)- 2 m^2 x^2\right\}\nonumber\\
&\times\theta\left(x^{q\to qg}_+-x\right)\theta\left(x-x^{q\to qg}_-\right)  \,,\label{C11a}\\
\frac{dW_{g_{\lambda_0}\rightarrow gg}}{d x}=&\frac{3\alpha_s g^2}{4x^2 (1-x)^2E }\left\{\left(b_0 E x-\omega_p^2\right) (1-x)-\omega_p^2x^2\right\}  \nonumber \\
&\times\left\{ \left[x^4+(1-x)^4\right]\delta_{\lambda_0,-1}+\delta_{\lambda_0,1} \right\}\nonumber\\
&\times\theta\left(x^{g\to gg}_+-x\right)\theta\left(x-x^{g\to gg}_-\right)
\,,\label{C11d}
\gal
\end{subequations}
where $\lambda_0=\pm 1$ is the right/left-hand polarizations of the incident gluon.

\begin{figure}[ht]
\begin{tabular}{cc}
      \includegraphics[width=0.5\linewidth]{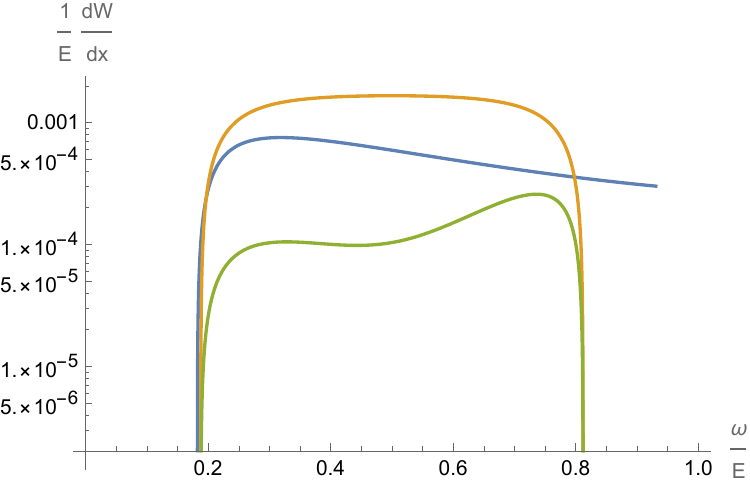} &
      \includegraphics[width=0.5\linewidth]{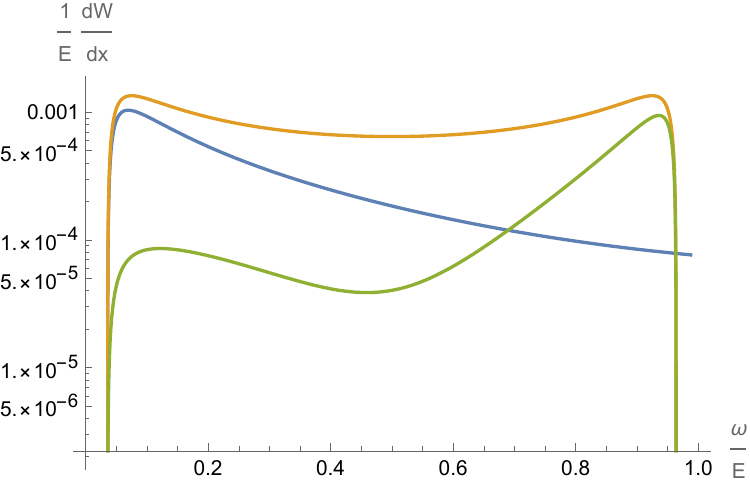}  
      \end{tabular}
  \caption{The Color Chiral Cherenkov radiation rates versus $x=\omega/E$. 
  The channels $q\rightarrow q g$, $g_R\rightarrow g g$, and $g_L\rightarrow g g$ are represented by blue, orange, and green lines respectively. Left panel: $E=20$~GeV,  right panel: $E=100$~GeV. Both panels: $g=2$, $T=300$~MeV, $b_0=50$~MeV.}
\label{fig:rateplot}
\end{figure}

We now consider the rate of gluon radiation due to the color chiral magnetic current in quark-gluon plasma. The plasma frequency and the quark thermal mass are given by $\omega_p^2=\frac{g^2T^2}{18}(2N_c+N_f)$ and  $m^2=\frac{g^2T^2}{16N_c}(N_c^2-1)$ respectively \cite{RISCHKE2004197}. Using these equations in \eq{C11} we compute the anomalous contribution to the gluon emission spectra from quark-gluon plasma shown in \fig{fig:rateplot}. We emphasize that this calculation takes into account only the anomaly--induced radiation which is but a fraction of the total gluon radiation.  
One can observe that the gluon spectrum is constrained by the thresholds \eq{t13},\eq{t19}. The thresholds nearly coincide in the infrared, but are different in the ultraviolet part of the spectrum. The right-handed gluons decay most readily except in the far infrared which is dominated by the radiation off the quark.

\begin{figure}[ht]
\begin{tabular}{cc}
      \includegraphics[width=0.7\linewidth]{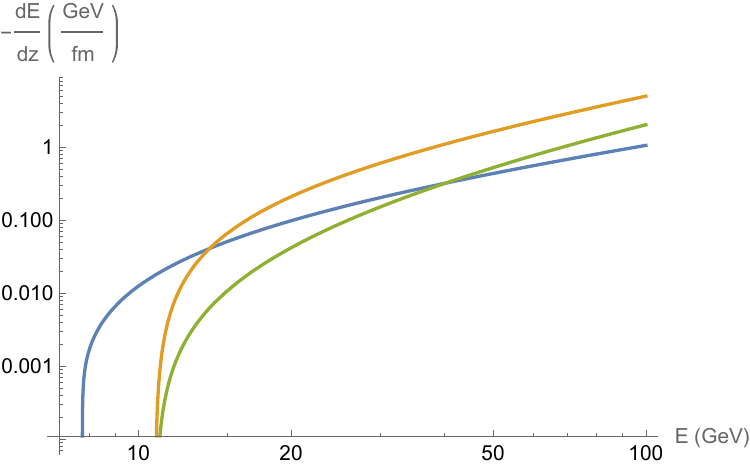}   
      \end{tabular}
  \caption{The rate of energy loss due to the Color Chiral Cherenkov radiation for $q\rightarrow q g$ (blue), $g_R\rightarrow g g$ (orange), and $g_L\rightarrow g g$ (green). Parameters: $g=2$, $T=300$~MeV, $b_0=50$~MeV.  }
\label{fig:loss}
\end{figure}

The contribution of the Color Chiral Cherenkov radiation to the energy loss by a fast parton propagating in quark-gluon plasma along the $z$-axis is given by 
\ball{t20}
-\frac{dE_{a\to bc}}{dz} = E\int_0^1 x\frac{dW_{a\to bc}}{dx}dx\,.
\gal
Substituting \eq{C11} into \eq{t20} yields for each channel we obtain in the high energy limit $b_0 E\gg m^2,\omega_p^2$:
\begin{subequations}\label{C9b}
\bal
&-\frac{dE_{q\rightarrow qg}}{d z}=\frac{4\alpha_s g^2b_0E}{ 9}\,,\label{C9ba}\\
&-\frac{dE_{g_R\rightarrow gg}}{d z}=\frac{3\alpha_s g^2b_0 E}{4 }\left(\ln{\frac{b_0 E}{\omega_p^2}}-1\right) \,,\label{C9bb}\\
&-\frac{dE_{g_L\rightarrow gg}}{d z} =\frac{3\alpha_s g^2b_0 E}{4}\left(\ln{\frac{b_0 E}{\omega_p^2}}-\frac{17}{6}\right)\,.\label{C9bc}
\gal
\end{subequations}
\fig{fig:loss} exhibits the contribution of the Color Chiral Cherenkov radiation to the parton energy loss.  We observe in that at high energy the right-hand gluon loses more energy than the left-handed one and the quark, while at lower energy the quark channel is the main mechanism of energy loss. 

Comparing the decay rates of the left-handed and right-handed gluons in \fig{fig:rateplot}, one finds that in a medium with $b_0>0$, the left-handed gluons decay more slowly when compared to right-handed ones. As a result jets develop strong left-hand polarziation. This is the clearest manifestation of the chiral anomaly in the jet structure. 
 
 The chiral imbalance is also imprinted into the jet loss pattern seen in  \fig{fig:loss}: the right-handed gluons loose a lot more energy than the left-handed ones.  However, as the jet energy increases, the difference between the gluon polarizations becomes less pronounced since the energy loss is driven primarily by the large polarization independent logarithms in \eq{C9b}. These logarithms also enhance the relative contribution of gluons as compared to quarks to the energy loss at high energies. In contrast, at low energy, the energy loss is dominated by quarks as seen in \fig{fig:loss}. We stress again that these conclusions hold only for energy loss due to the Color Chiral Cherenkov radiation, ignored all other contributions. A recent review of the conventional mechanisms of energy loss can be found in \cite{Cao_2021}. 
 
In conclusion, energy loss due to the Color Chiral Cherenkov radiation is significant. Therefore a comprehensive phenomenological analysis requires incorporation of the novel energy loss channels into the numerical packages describing jets in hot nuclear medium.

\section{Bremsstrahlung in chiral medium}\label{sec:bremss}


Bremsstrahlung dominates the energy loss in conventional media at high energies. This section deals with the novel contributions to bremsstrahlung that arise in chiral media. In additional to the modification of the gauge boson dispersion relations that is the cause of the Chiral Cherenkov radiation which we discussed in the previous sections, the chiral contributions to bremsstrahlung come from the chiral terms of the gauge boson propagator. In this section we do the calculations in the framework of the axion QED.  Most results can be readily generalized to the strong interactions. We divided this section in two parts dealing with the electric and magnetic terms in the modified photon propagator respectively \cite{Hansen:2023wzp,Hansen:2022nbs}.

\subsection{Photon propagator and gauge symmetry in  chiral medium}\label{Vsec:app1}

The classical equation of motion of the photon field is 
\ball{Vap1}
\left[ g^{\mu\nu}\partial^2-\partial^\mu\partial^\nu-\epsilon^{\mu\nu\alpha\beta}b_\alpha\partial_\beta\right] A_\nu= j^\mu\,.
\gal
It is invariant under the gauge transform 
\ball{Vap2}
A_\nu\to A_\nu +\partial_\nu\chi\,
\gal
where $\chi$ is an arbitrary function. In the Lorenz gauge $\partial\cdot A=0$ the photon Green's function obeys the equation
\ball{Vap3}
\left[ g^{\mu\nu}\partial^2-(1-1/\xi)\partial^\mu\partial^\nu-\epsilon^{\mu\nu\alpha\beta}b_\alpha\partial_\beta\right] D_{\nu\lambda}(x)= i\delta\indices{^\mu_\lambda} \delta^4(x)\,,
\gal
where $\xi$ is the gauge parameter. In momentum space \eq{Vap3} becomes
\ball{Vap5}
\left[ -k^2g^{\mu\nu}-(1-1/\xi)k^\mu k^\nu+i\epsilon^{\mu\nu\alpha\beta}b_\alpha k_\beta\right] D_{\nu\lambda}(k)= i\delta\indices{^\mu_\lambda}\,,
\gal
which is solved by\cite{Hansen:2022nbs}
\ball{Vap7}
 D^{\nu\lambda}(k)= &-\frac{i}{k^4+b^2k^2-(k\cdot b)^2}
\bigg\{ k^2 g^{\nu\lambda}+b^\nu b^\lambda+i\epsilon_{\nu\lambda\alpha\beta}b^\alpha k^\beta
\nonumber\\
&- \frac{(b\cdot k)}{k^2}(k^\nu b^\lambda+k^\lambda b^\nu)
+\left[b^2\xi-(1-\xi)\left( k^2-\frac{(b\cdot k)^2}{k^2}\right)\right]\frac{k^\lambda k^\nu}{k^2}
\bigg\}
\gal
The terms in the second line vanish when $D^{\nu\lambda}$ is inserted in the Feynman diagram due to the current conservation. Similar expression was obtained in \cite{Adam:2001ma}.


The wavefunction of a photon with given momentum and polarization at finite  $b=(\sigma_\chi,\b 0)$ is a plane wave 
\ball{Vap20}
A^\mu(x) =\frac{1}{\sqrt{2\omega}}e^\mu_k e^{-ik\cdot x}\,,
\gal
where $e^\mu_k= (0,\b e_k)$ is a circular polarization vector satisfying the 
 the Lorenz gauge $k\cdot e_k=\b k\cdot \b e_k=0$. Once the wave functions \eq{Vap20} are chosen, there is no remaining gauge freedom. Only circularly polarized photons solve the equation of motion \eq{Vap1}. 
 This stands in stark contrast with the free photon wavefunction which is also given by the plane wave as in \eq{Vap20}. However, since $k^2=0$, there still remains gauge freedom to transform the polarization vector  $e^\mu\to e^\mu +\chi k^\mu$ without changing the form of the wavefunction or violating the Lorenz gauge condition. 
 
The photon polarization sum is
\ball{Vap23}
d^{ij}= \sum_\text{pol}e^i_ke^{*j}_k=\delta^{ij}-\frac{k^ik^j}{\b k^2}\,.
\gal 
There are no remaining gauge freedom to transform it to any other form. In particular $d^{ij}$ cannot be replaced with $-g^{\mu\nu}$. As an illustration consider the amplitude $e_k\cdot \mathcal{M}$. Let $k$ be in the $z$-direction, then using the current conservation $k\cdot \mathcal{M}=0$ we can write
$$
\sum_\text{pol}|e_k\cdot \mathcal{M}|^2=|\mathcal{M}_x|^2+|\mathcal{M}_y|^2= |\mathcal{M}_x|^2+|\mathcal{M}_y|^2+|\mathcal{M}_z|^2-\frac{\omega^2}{\b k^2}|\mathcal{M}_0|^2\neq -g^{\mu\nu}\mathcal{M}_\mu\mathcal{M}^*_\nu\
$$
since $k^2= \omega^2-\b k^2\neq 0$.

\subsection{Electric monopole (Coulomb) scattering}\label{Wsec:electric}


Following the original Bethe and Heitler calculation \cite{Bethe:1934za}, we consider scattering of a charged fermion off a heavy nucleus carrying electric charge $eZ$ and magnetic moment $\textgoth{m}$. The differential cross section for photon radiation is given by the familiar expression
\ball{Wc1}
    d\sigma=\frac{1}{2}\sum_{ss'\lambda}\left|\mathcal{M}\right|^2
    \frac{1}{8(2\pi)^5}\frac{\omega |\b p'|}{|\b p|} d\Omega_{\bm{k}} d\Omega' d\omega\,,
\gal
where  $p=(E,\b p)$, $p'=(E',\b p')$ and $k=(\omega,\b k)$ are the incoming fermion, outgoing fermion, and radiated photon momenta respectively. 
The matrix element $\mathcal{M}$ reads:
\ball{Wc3}
    \mathcal{M}=e^2 \overline{u}(p')\left(\slashed{e}_{k\lambda}^*\frac{ \slashed{p}'+\slashed{k}+m}{2p'\cdot k+k^2}\slashed{A}(\b q)-\slashed{A} (\bm{q})\frac{ \slashed{p}-\slashed{k}+m}{2p\cdot k-k^2}\slashed{e}_{k\lambda}^*\right)u(p)\,,
\gal
where $A(\b q)$ indicates the external field, $q=p'-p+k$ is the momentum transfer, $m$ is the mass of the fermion, and $e_{k\lambda}$ is the photon polarization vector. Whereas in vacuum the photon four-momentum is light-like, in chiral media it is not, as is evident from \eq{Xi1}. Moreover, in the Lorentz gauge, photon is forced to be in one of the two  circularly polarized states. This eliminates the residual gauge invariance that is reflected in the Ward identity \cite{Hansen:2022nbs}.

The electromagnetic potential induced by the electric current $J^\mu$, associated with the nucleus, in a chiral medium is $A_\mu=-iD_{\mu\nu}J^\nu$. The photon propagator in chiral medium in the Lorentz/Landau gauge take form \cite{Hansen:2022nbs}: 
\ball{Wa5}
D_{\mu\nu}(q)=& -i \frac{q^2 g_{\mu\nu}+i\epsilon_{\mu\nu\rho \sigma}b^\rho q^\sigma+b_\mu b_\nu}{q^4+b^2 q^2-(b\cdot q)^2}\nonumber\\
&+i\frac{\left[q^2-(b\cdot q)^2/q^2\right]q_\mu q_\nu + b\cdot q(b_\mu q_\nu+b_\nu q_\mu)}{q^2\left[q^4+b^2 q^2-(b\cdot q)^2\right]}\,.
\gal
In the static limit $q^0=0$ the components of the photon propagator read \cite{Qiu:2016hzd}
\begin{subequations}\label{Wb12}
\bal
&D_{00}(\b q)= \frac{i}{\b q^2}\,,\label{Wb12a}\\
&D_{0i}(\b q)= D_{0i}(\b q)= 0\,,\label{Wb12b}\\
&D_{ij}(\b q)=-\frac{i\delta_{ij}}{\b q^2-\sigma_\chi^2}-\frac{\epsilon_{ijk}q^k}{\sigma_\chi(\b q^2-\sigma_\chi^2)}+\frac{\epsilon_{ijk}q^k}{\sigma_\chi\b q^2}+\frac{iq_iq_j}{\b q^2(\b q^2-\sigma_\chi^2)}\,.\label{Wb12c}
\gal
\end{subequations}
The gauge-dependent terms proportional to $q_\mu$ and $q_\nu$ vanish when substituted into the scattering amplitude. The spatial components \eq{Wb12c} couple to the nucleus magnetic moment and have a resonance at $\b q^2=\sigma_\chi^2$. This resonant term couples to the nucleus magnetic moment $\textgoth{m}$. We consider in the next subsection. Here we are interested in the electric monopole component of the external field which is determined by the nuclear electric charge. Convolution of the current $J^\nu(\b x) =eZ \delta\indices{^\nu_0}\delta(\b x)$ with the $D_{00}$ component of the photon propagator gives rise to the Coulomb potential:
\ball{Wb15}
A^0(\b q)= eZ/\b q^2\,,\qquad \b A(\b q)=0\,.
\gal
Plugging  \eq{Wb15} into \eq{Wc3} and averaging over the fermion spin directions, we can obtain an expression for the  differential cross section. The reader is referred to \cite{Hansen:2023wzp} for the explicit expression.  

An examination of the fermion and photon propagators used in the calculation of the differential cross section reveals divergences in three kinematic regions: $\b q^2=0$, $2\omega\kappa=k^2$ and  $2\omega \kappa'=-k^2$, where $\kappa=p\cdot k/\omega$ and $\kappa'=p'\cdot k/\omega$ The first one is the familiar infrared Coulomb pole $\b q^2=0$ which is regulated in the usual way  \cite{Wang:1994fx,Baier:1996vi}:
\ball{Wf2}
\frac{1}{\b q^2}\to \frac{1}{\b q^2+\mu^2}\,,
\gal
where $\mu$ is the Debye mass of the medium. If the  scattering particle mass $m$ is much larger than $\mu$, the minimum momentum transfer is of the order $m$. However, the Debye mass scales with the temperature and becomes much larger than $m$ in a sufficiently hot medium.\footnote{In hot plasmas the long-distance terms neglected in \eq{f2}, may become important\cite{Laine:2006ns}.}  Therefore, in the following analysis it will be convenient to consider two limiting cases depending on the relative magnitude of $m$ and $\mu$.  

The other two divergences occur in the fermion propagator and  reflect its instability in chiral matter with respect to spontaneous photon emission  \cite{Carroll:1989vb,Joyce:1997uy,Boyarsky:2011uy,Kharzeev:2013ffa,Khaidukov:2013sja,Kirilin:2013fqa,Akamatsu:2013pjd,Avdoshkin:2014gpa,Dvornikov:2014uza,Tuchin:2014iua,Manuel:2015zpa,Buividovich:2015jfa,Sigl:2015xva,Xia:2016any,Kaplan:2016drz,Kirilin:2017tdh,Tuchin:2018sqe,Mace:2019cqo}. With the account of the finite width, the fermion propagators are modified as follows:  
\ball{Wf1}
\frac{1}{2\omega\kappa-k^2}\to \frac{1}{2\omega\kappa-k^2+iE/\tau}\,,\qquad \frac{1}{2\omega\kappa'+k^2}\to \frac{1}{2\omega\kappa'+k^2-iE'/\tau}\,,
\gal
where $\tau$ is the relaxation time as measured in the medium rest frame. In the fermion rest frame the relaxation rate is $(E/m)\tau^{-1}$. A number of inelastic processes contribute to the relaxation of the chiral state of the fermion. Among them is the Chiral Cherenkov radiation whose rate is given by \eq{Xd55} and whose spectrum stretches up to the cutoff energy $\omega^*$ given by \eq{Xd52}:
\ball{Wf5}
\omega^*=\frac{\lambda \sigma_\chi E^2}{\lambda \sigma_\chi E+m^2}\,.
\gal
In particular, when $m^2\gg \lambda \sigma_\chi E$, the fermion decay rate is $W\approx \alpha \sigma_\chi/2$. One can take this as the low bound of the total relaxation rate: $\tau^{-1}> \alpha \sigma_\chi/2$.  We can use \eq{Wf5} to eliminate $\lambda \sigma_\chi$ in favor of $\omega^*$:
\ball{Wj2.5}
-k^2\approx \lambda \sigma_\chi\omega = \frac{\omega\omega^* m^2}{E(E-\omega^*)}\,,\quad (\lambda \sigma_\chi>0)\,.
\gal
Note that \eq{Wj2.5} makes sense only when $\lambda \sigma_\chi>0$, for otherwise $\omega^*$ is negative, indicating that there is no instability when $\lambda \sigma_\chi<0$.

Let us now examine the photon propagator in the resonant case $\lambda b_0>0$. Writing $\b q^2= [(\b k\times \b q)^2+ (\b k\cdot \b q)^2]/\b k^2$ and expanding at small photon emission angles we obtain 
\ball{Wj3}
\b q^2=-(k+p'-p)^2\approx  \theta^2E^2+\theta'^2E'^2-2EE'\theta \theta'\cos\phi\nonumber\\
+\frac{1}{4}\left[\frac{m^2(\omega-\omega^*)}{E'(E-\omega^*)}-E \theta^2+E' \theta'^2\right]^2\,.
\gal
In the non-anomalous case $\omega^*=0$, and the momentum transfer is bounded from below by $\frac{m^4\omega^2}{4 E^2E'^2}$. In contrast, in the presence of the anomaly the momentum transfer is allowed to vanish. We can find the corresponding kinematic region by first observing that the sum of the first three terms and the last term in the r.h.s.\ of \eq{Wj3} are non-negative and therefore have to vanish independently. The sum of the first three terms vanishes only when $\phi=0$ and $E \theta=E' \theta'$ in which case the momentum transfer reads
\ball{Wj4}
\b q^2|_{\phi=0,E \theta=E' \theta'}\approx \frac{1}{4}\left[\frac{m^2(\omega-\omega^*)}{E'(E-\omega^*)}+\frac{\omega E}{E'} \theta^2\right]^2 =
\frac{1}{4}\frac{\omega^2 E^2}{E'^2}\left[\frac{m^2(\omega-\omega^*)}{\omega E(E-\omega^*)}+\theta^2\right]^2\,.
\gal
This imbues the photon propagator with the same resonant behavior as the fermion propagator. Apparently, we need to regulate the divergence at $\b q^2=0$ by replacing  $\theta^2\to \theta^2+\frac{i}{\omega \tau}$ in \eq{Wj4}. This is tantamount to the replacement $\b q^2\rightarrow\b q^2+\frac{E^2}{4E'^2\tau^2}$. Along with the Debye mass $\mu$ introduced in \eq{Wf2} it provides the regulator of the photon propagator at small momentum transfers: 
\ball{Wj10}
\b q^2\rightarrow\b q^2+\frac{E^2}{4E'^2\tau^2}+\mu^2\,.
\gal

In applications one is usually interested in the high energy limit $ E, E'\gg m,\mu$ where
the expression for the differential cross section and squared matrix element significantly simplify. Moreover, the physics of bremsstrahlung in chiral medium is most transparent in two limiting cases: (i) low temperature $m\gg \mu$ and (ii) high temperature $\mu\gg m$. In the anomaly-free medium  $\sigma_\chi=0$, the first case reduces to the scattering off a single nucleus. In this case the momentum transfer $\b q^2$ never vanishes. On the contrary, in the presence of chiral anomaly, negative $k^2$ can drive the momentum transfer towards zero for one of the photon polarizations. 

\subsubsection{Low temperature $\mu\ll m$}\label{Wsmall mu}

We first consider the low temperature/heavy fermion regime. We also assume, for the sake of simplicity, that $\mu^2\ll 1/\tau^2$ so that $\tau$ regulates both the fermion and the photon propagators. 
In the anomaly-free medium, the typical momentum transfer is of the order $m$ and therefore the scattering cross section is insensitive to the cutoffs $\tau$ and $\mu$. This conclusion is upended in the anomalous medium.

The bremsstrahlung cross section is essentially different for positive and negative values of the parameter $\sigma_\chi\lambda$. For $\sigma_\chi\lambda<0$ there are no resonances and the angular integrals simplify greatly. In this case, $\tau^{-1}$ and $\mu$ may be neglected given that all integrals are convergent. In this case the cross section reads\cite{Hansen:2023wzp}: 
\ball{Wsleft}
    \frac{d\sigma(\sigma_\chi\lambda<0)}{d\omega}\approx \frac{Z^2e^6 E'}{4(2\pi)^3(m^2\omega-\lambda \sigma_\chi  E E') E}
    \left(\frac{ E}{ E'}+\frac{ E'}{ E}-\frac{2}{3}\right)\nonumber\\
    \times\left[\ln\frac{4 E^2 E'^2}{\omega(m^2\omega-\lambda \sigma_\chi  E E')}-1\right]\,.
\gal
In the anomaly-free medium $\sigma_\chi= 0$  \eq{Wsleft} reduces to the well-known Bethe-Heitler expression for the bremsstrahlung cross section on a heavy nucleus
\cite{Bethe:1934za,Berestetskii:1982qgu}:
\ball{Wk5}
    \frac{d\sigma_\text{BH}}{d\omega}\approx \frac{ Z^2e^6 E'}{4(2\pi)^3m^2\omega E}\left(\frac{ E}{ E'}+\frac{ E'}{ E}-\frac{2}{3}\right)\left(\ln\frac{2 E E'}{m\omega}-\frac{1}{2}\right)\,.
\gal
The effect of the anomaly on this photon polarization  is most significant in the infrared region $\omega\ll \sigma_\chi EE'/m^2$ where the cross section scales as $\log(1/\omega)$. In contrast, without the anomaly it scales as $(1/\omega)\log(1/\omega)$.
Thus the anomaly tends to suppress emission of photons with $\sigma_\chi\lambda<0$ polarization. 

The situation is essentially different for the $\sigma_\chi\lambda>0$ polarziation since the corresponding cross section diverges in the limit $\tau^{-1}\to 0$, i.e.\ at small photon emission angles when $\omega\le \omega^*$. Careful analysis reveals that this divergence occurs concurrently in the fermion and the photon propagators and is regulated by shifting $\theta^2\to \theta^2+\frac{i}{\omega \tau}$, where $\theta$ is the emission angle. The final result in the limit $\mu=0$ reads\cite{Hansen:2023wzp}:
\ball{Wsright2}
    \frac{d\sigma (\sigma_\chi\lambda>0)}{d\omega}\approx \frac{Z^2e^6 E'}{4(2\pi)^3 E m^2\omega}
    \bigg\{
    \left(\frac{E^2(\omega^*-\omega)^2}{\omega^2(E-\omega^*)^2}+\frac{4 E^4}{m^4\omega^2\tau^2}\right)^{-1/2}
        \nonumber\\  
   \times\left(\frac{ E}{ E'}+\frac{ E'}{ E}-\frac{2}{3}\right) \left(\ln\frac{4 E^2 E'^2}{m^2\omega^2\sqrt{\frac{E^2(\omega^*-\omega)^2}{\omega^2(E-\omega^*)^2}+\frac{4 E^4 E'^2}{m^4\omega^4\tau^2}}}-1\right)
    \nonumber\\
     +\frac{2m^4 (\omega^*-\omega)\tau^3}{E^3E'^4  (E-\omega^*)}\left[E'^6\arctan\frac{m^2(\omega^*-\omega)\tau}{  E'(E-\omega^*)} +E^6\arctan\frac{m^2(\omega^*-\omega)\tau }{ E (E-\omega^*)}\right]
     \nonumber\\
     \times\theta(\omega^*-\omega)\bigg\}\,,
\gal
 where we replaced $\sigma_\chi\lambda$ in favor of  $\omega^*$  using \eq{Wj2.5}. 

There are two kinds of terms in \eq{Wsright2}. (i) The first line of \eq{Wsright2} reduces to the anomaly-free result \eq{Wk5} in the limit $\sigma_\chi\to 0$ and $ E/\tau\to 0$. We neglected all terms proportional to $\tau^{-1}$ with exception of those appearing under the radicals where their role is to regulate the divergence at the threshold $\omega=\omega^*$. 
(ii) The second and the third lines of \eq{Wsright2} represent the most singular resonance contributions, viz.\ the terms that are most divergent at small $\tau^{-1}$. In the limit $\sigma_\chi\to 0$ the step function can only be satisfied when $\omega\to 0$, hence the anomalous contribution vanishes.

 We stress that the  resonance in the fermion propagator contributes only to one of the photon polarizations, namely $\sigma_\chi\lambda>0$. The other polarization $\sigma_\chi\lambda<0$ is suppressed. This is in contrast to the lack of a distinction between handedness in the absence of the anomaly. A remarkable feature of the anomalous contribution, dominated by the second and the third lines of \eq{Wsright2}, is that at small $\omega$ its spectrum scales as $1/\omega$, which is the same form as the soft photon spectrum without the chiral anomaly \eq{Wk5}. 
 
 \begin{figure}[t]
\begin{tabular}{cc}
      \includegraphics[height=4.cm]{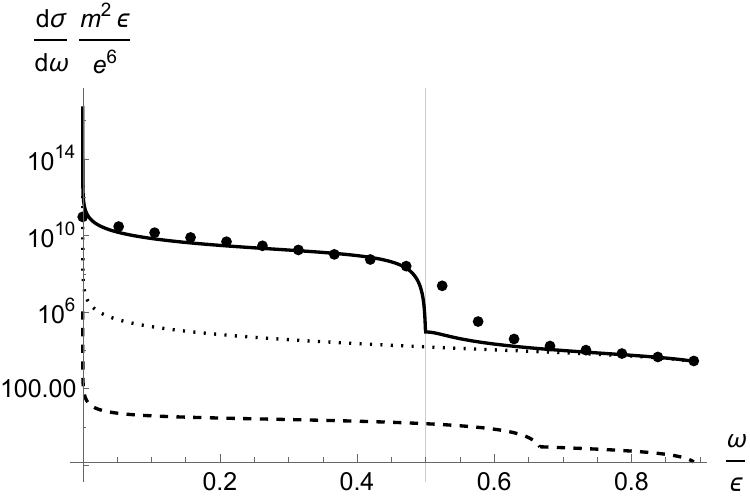} &
      \includegraphics[height=4.cm]{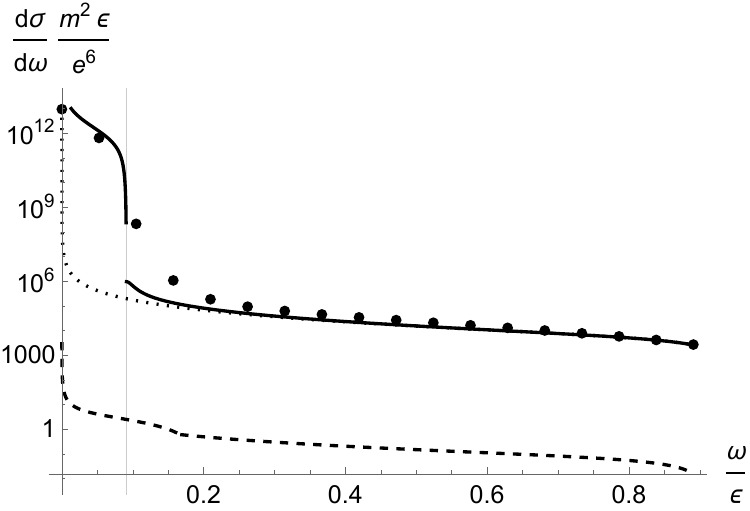}
      \end{tabular}
  \caption{The bremsstrahlung spectrum at $\mu\ll m$. 
  The dots: the exact leading order, solid line: the high energy approximation given by the sum of equations \eq{Wsleft} and \eq{Wsright2}, dotted line: the Bethe-Heitler expression \eq{Wk5}, dashed line: the magnetic moment contribution (see the next section). 
Left panel:  $\sigma_\chi=0.1m$,  $\mu=10^{-3}\sigma_\chi$, $\tau^{-1}= 0.1\sigma_\chi$. Right panel: $\sigma_\chi=10^{-2}m$, $\mu=10^{-3}\sigma_\chi$, $\tau^{-1}=0.1 \sigma_\chi$. Both panels: $E=10m$,  $Z=33$, $\textgoth{m}=\mu_N$, $\Gamma=0.1\sigma_\chi$. The vertical line indicates $\omega^*/E$.}
\label{Wfig:spectrum}
\end{figure}
 
 The bremsstrahlung photon spectra are plotted in \fig{Wfig:spectrum}.  Inspection of \fig{Wfig:spectrum} reveals two significant features. Firstly, at $\omega<\omega^*$ the cross section is enhanced as compared to the Bethe-Heitler formula.

 
Let us now compute the amount of energy lost due to the bremsstrahlung in a scattering off a single source of electric charge $eZ$.\footnote{A comprehensive discussions of the energy loss regimes and phenomenological applications can be found in \cite{Workman:2022ynf} and \cite{Peigne:2008wu}.}
 Throughout this section we neglected the quantum interference effects assuming that the photon formation time $t_f$ is much shorter than the mean-free-path $\ell$. In this approximation  
the energy loss per unit length is:
\begin{align}\label{We.loss.1}
-\frac{d E}{d z}= n\int_0^E\omega \frac{d\sigma^{eZ\to eZ\gamma}}{d\omega}d\omega \,,
\end{align}
where $n$ is number of scatterers per unit volume which can be expressed in terms of the elastic scattering cross section $n=1/\ell\sigma^{eZ\to eZ}$. At high energies, 
\bal\label{We.loss.2}
\sigma^{eZ\to eZ}= \frac{\alpha (eZ)^2}{\max\{\mu^2,m^2\}}+\mathcal{O}(\textgoth{m})\,,
\gal
where $\alpha= e^2/4\pi$.
The omitted term in \eq{We.loss.2} is proportional to the small magnetic moment $\textgoth{m}$ of the source. 
We discuss it in the next subsection. 

Substituting \eq{Wsleft} or \eq{Wsright2} into \eq{We.loss.1} and integrating over $\omega$ gives the energy 
loss for two photon polarizations.  For brevity we will record these expressions in the limit $m^2\gg \sigma_\chi E$:
\bal\label{We.loss.5}
  -\frac{d E(\sigma_\chi\lambda<0)}{d z}\approx \frac{e^2 E}{16\pi^2\ell}\left[\ln\frac{2 E}{m}-\frac{1}{3}-\frac{2\sigma_\chi  E}{9m^2}\left(\pi+2\ln\frac{ E}{\sigma_\chi}\right)\right]\,,
\gal
\bal\label{We.loss.6}
 -\frac{d E(\sigma_\chi\lambda>0)}{d z}\approx  \frac{e^2 E}{16\pi^2\ell}
    \Bigg\{\ln\frac{2 E}{m}-\frac{1}{3}+\frac{2\sigma_\chi  E}{9m^2}\left(\pi+2\ln\frac{ E}{\sigma_\chi}\right)\nonumber\\
     +2\tau   (E-\omega^*)\arctan\frac{2m^2\omega^*\tau}{  E(E-\omega^*)}\Bigg\}\,.
\gal
Eq.~\eq{We.loss.5} agrees with Eq.~(93.24) in \cite{Berestetskii:1982qgu} when $\sigma_\chi=0$. We observe that the anomaly slightly reduces the amount of energy lost due to the bremsstrahlung of $\sigma_\chi\lambda<0$ photons. This  is consistent with our discussion in the previous section that the bremsstrahlung cross section is reduced in this case.

In summary, we observe in \fig{Wfig:spectrum} that at low temperatures $\mu\ll m$ the anomalous contribution to bremsstrahlung, stemming from one of the photon polarizations, is several orders of magnitude larger than the non-anomalous one. If confirmed by experimental observation, it can serve as an effective tool to study the chiral anomaly in the materials. It can also be used to search for the new forms of the chiral matter. In particular, a cosmic ray moving through the chiral domain generated by an axion field would radiate and lose energy in a peculiar way described in this subsection.

\subsubsection{High temperatures $\mu\gg m$}\label{Whigh T}

We now consider the differential cross-section in the opposite limit where $\mu\gg m$ and $\mu^2\gg 1/\tau^2$, so that $\tau^{-1}$ is neglected in the photon propagator.  We assume that the projectile fermion experiences only a single hard scattering and therefore its propagator is the same as in vacuum \cite{Wang:1994fx,Baier:1996vi}. The result of a lengthy calculation is different for the two photon polarziations and reads\cite{Hansen:2023wzp} 
\ball{Wg5}
    \frac{d\sigma(\sigma_\chi\lambda<0)}{d\omega}\approx \frac{Z^2e^6 E'}{4(2\pi)^3(\mu^2\omega -\lambda \sigma_\chi  E E') E}
    \left(\frac{ E}{ E'}+\frac{ E'}{ E}\right)\left[\ln\frac{4 E^2 E'^2}{\mu^2\omega^2-\lambda \sigma_\chi \omega E E'}-1\right]\,,
\gal
\ball{Wg7}
    &\frac{d\sigma (\sigma_\chi\lambda>0)}{d\omega}\approx \frac{Z^2 E'e^6}{4(2\pi)^3 E\mu^2\omega}
    \Bigg\{ \frac{\mu^2\omega(\frac{ E}{ E'}+\frac{ E'}{ E})\big[\ln\frac{4 E^2 E'^2}{\sqrt{(\mu^2\omega^2-\lambda \sigma_\chi \omega E E')^2+4 \frac{E^4E'^2}{\tau^2}}}-1\big]}{\sqrt{(\mu^2\omega-\lambda \sigma_\chi  E E')^2+ \frac{E^3E'}{\tau^2}}}
    \nonumber\\
     &  +\frac{\lambda  \sigma_\chi\mu\tau }{  E} \arctan\left(\frac{(\lambda \sigma_\chi  E E'-\mu^2\omega)\tau}{ E E'}\right)\theta(\omega^*-\omega)\bigg\}\,,
\gal
where $\omega^*$ is given by \eq{Wf5}  and $ E', E\gg \mu$.

\begin{figure}[t]
\begin{tabular}{cc}
      \includegraphics[height=4.4cm]{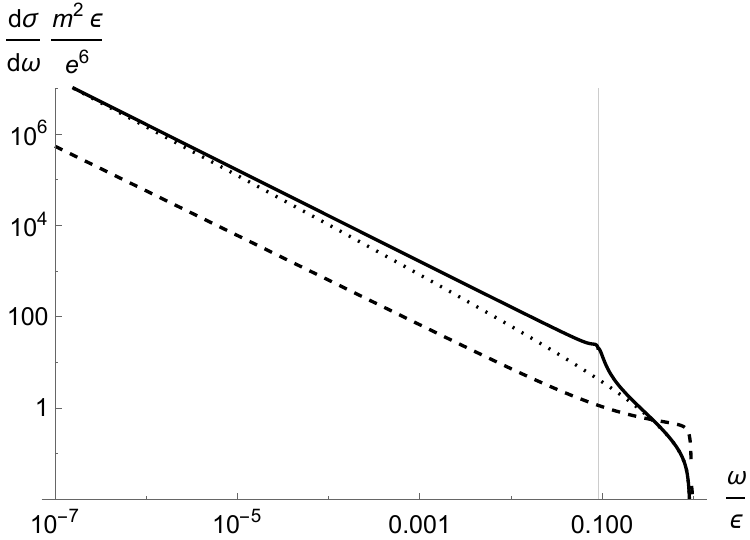} &
      \includegraphics[height=4.4cm]{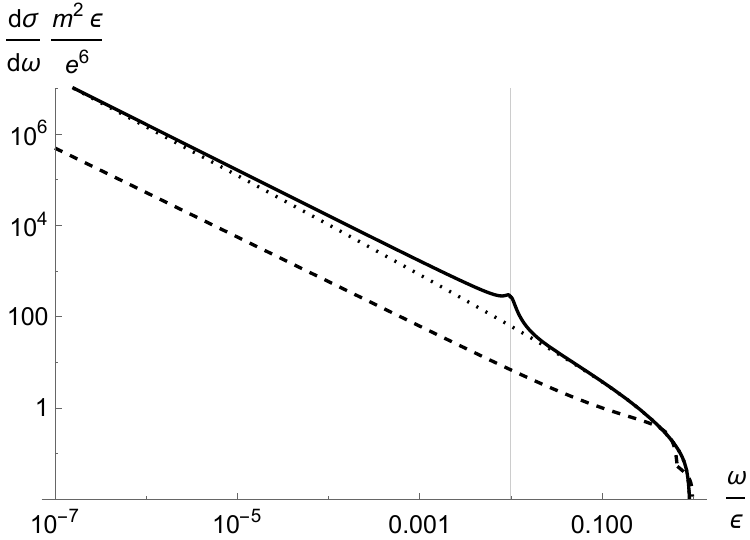} 
      \end{tabular}
  \caption{ The bremsstrahlung spectrum at $\mu\gg m$. Solid line: the high energy approximation given by the sum of equations \eq{Wg5} and \eq{Wg7}, dotted line: the Bethe-Heitler limit $\sigma_\chi=0$, dashed line: the magnetic moment contribution. Left panel: 
  $\sigma_\chi=10^{-1} m$, right panel: $\sigma_\chi=10^{-2}m$, both panels: $E=10^2 m$,  $\mu=10 m$, $\tau^{-1}=\Gamma=0.1 \sigma_\chi$, $Z=33$. The vertical line indicates $\omega^*/E$.}
\label{Wfig:spectrum2}
\end{figure}

The rate of energy loss in the high-temperature limit is computed similarly. Plugging \eq{Wg5} and \eq{Wg7} into \eq{We.loss.1} we derive
\bal\label{We.loss.14}
    -\frac{d E(\sigma_\chi\lambda<0)}{d z}&\approx \frac{e^2 E}{16 \pi^2 \ell}\left(\ln\frac{2 E}{\mu}-\frac{6\sigma_\chi  E}{7\mu^2}\ln\frac{ E}{\sigma_\chi}\ln\frac{4 E^2}{\mu^2}\right)\,,\\
\label{We.loss.15}
    -\frac{d E(\sigma_\chi\lambda>0)}{d z}&\approx \frac{e^2 E}{16 \pi^2 \ell}\left(\ln\frac{2 E}{\mu}+\frac{6\sigma_\chi  E}{7\mu^2}\ln\frac{ E}{\sigma_\chi}\ln\frac{4 E^2}{\mu^2}+\frac{ 2 \sigma_\chi \mu\tau}{ 3 E}\right)\,.
\gal
Eq.~\eq{We.loss.14} agrees with \cite{Baier:1994bd} when $\sigma_\chi=0$. As in the  low temperature case, the energy loss is mostly driven by $\sigma_\chi \lambda>0$ polarization. However, the overall magnitude of the lost energy sensitively depends on the actual values of the parameters.  

The total energy loss is given by  the sum of \eq{We.loss.5} and \eq{We.loss.5} (when $\mu\ll m)$  and of \eq{We.loss.14} and \eq{We.loss.15} (when $\mu\gg m$). We observe that the chirality dependent non-resonant contributions cancel out, while the resonant terms proportional to $\tau$ remain. Thus the total energy loss is also chirality dependent which may facilitate its experimental observation. 

In summary, at very high temperatures $\mu\gg m$, the effect of the anomaly is much smaller than ay low temperatures, as seen in \fig{Wfig:spectrum2}.


\subsubsection{Discussion}\label{Wsec:summary}

In this subsection we discussed  the bremsstrahlung radiation emitted by a fast particle  traveling in the chiral medium supporting the chiral magnetic current with constant chiral conductivity $\sigma_\chi$. In arriving at our conclusions we made a number of assumptions. First of all, we treated the  chiral conductivity $\sigma_\chi$ as a  constant. However, it does evolve in time albeit slowly. Its temporal evolution is characterized by the relaxation time $\tau$, which we assumed to be the large parameter in our calculation. The spatial extent of the domain can be neglected as long as it is larger than photon formation time $t_f$ which is a very good approximation considering that the formation time must be shorter than the mean free path $\ell$ in the Bethe-Heitler limit. 

Another critical assumption we made is that photon formation time $t_f$ is much shorter than the mean-free path $\ell$. As argued in \cite{Baier:1994bd}, the condition $t_f\ll \ell$ translates into the requirement that  $m,\mu\ll  E\ll \mu\sqrt{\omega\ell}$. Since $t_f$ is proportional to $\omega$, this approximation breaks down at high energies where one must take account of the multiple scatterings of the projectile in the medium. The resulting LPM quantum interference effect \cite{Landau:1953um,Migdal:1956tc} is an important feature of the bremsstrahlung spectrum and must certainly be taken into account at higher energies.  

\subsection{Magnetic dipole scattering}\label{Vsec:magnetic}

In the previous subsection we discussed scattering due to the Coulomb part $D_{00}$ of the photon propagator. Now we are interested to evaluate the contribution that comes about if the nucleus has magnetic moment $\textgoth{m}$. The magnetic moment couples to the $D_{ij}$ part of the photon propagator which has a resonance at $\b q^2= \sigma_\chi^2$, as evident in \eq{Wb12c}. 

In the point particle limit the spin current can be written as  $\b J(\b x) = \b \nabla \times (\textgoth{m}\delta(\b x))$. It represents the first non-vanishing  multipole moment of the vector potential. Altogether the electric current of the nucleus is
\ball{Vb16}
J^0(\b x) =eZ\delta(\b x)\,,\qquad 
\b J(\b x) = \b \nabla \times (\textgoth{m}\delta(\b x))\,,
\gal
which in momentum space reads
\ball{Vb17}
J^0(\b q)=eZ\,,\qquad \b J(\b q)= i\b q\times \textgoth{m}\,.
\gal
It  produces the vector potential 
\begin{subequations}\label{Vb19}
\bal
A^\ell(\b q)=&-iD^{\ell i}(\b q)J_i(\b q)\nonumber\\
=& -\frac{1}{\b q^2-\sigma_\chi^2}\left[ i(\textgoth{m}\times \b q)^\ell + \frac{\sigma_\chi}{\b q^2}(\textgoth{m}\cdot \b q q^\ell-\b q^2 \textgoth{m}^\ell)\right]\,. \label{Vb19b}
\gal
\end{subequations}
Eq.~\eq{Vb19b} satisfies the Coulomb gauge $\b q\cdot \b A=0$. The expression for this potential in the configuration space can be found in \cite{Tuchin:2020gtz}.

The differential cross section and the matrix element $\mathcal{M}$ are  given by \eq{Wc1} and   \eq{Wc3}  respectively. To simplify the calculation we restrict ourselves to the photon spectrum at $\omega\gg \sigma_\chi$. In this region, photon is approximately timelike up to the corrections of order $\sigma_\chi^2/\omega^2$. In other words, we neglect the anomaly in the photon wave function and thereby isolate the contribution of the resonance in the propagator. We also ignore the Debye mass in the present analysis. Since the direction of the nuclear magnetic moment $\textgoth{m}$ is random, we average over it using  $\aver{\textgoth{m}_i}=0$, $\aver{\textgoth{m}_i\textgoth{m}_j}=\frac{\textgoth{m}^2}{3}\delta_{ij}$ which yields
 \ball{Vc7}
     \left|\mathcal{M}\right|^2=\left|\mathcal{M}_e\right|^2+\left|\mathcal{M}_\mu\right|^2\,,
\gal
where $|\mathcal{M}_{e}|^2$ is proportional to $e^2Z^2$ and describes the term computed in \sec{Wsec:electric}. We now focus on the second term $|\mathcal{M}_\mu|^2$ which is proportional to $\textgoth{m}^2$.  Averaging over the magnetic moment directions in \eq{Vb19b} gives
\ball{Vc9}
\aver{A^i(\b q)A^{j*}(\b q)}=\frac{\mu^2}{3(\b q^2-\sigma_\chi^2)^2}\left[ \left(\delta^{ij}-\frac{q^iq^j}{\b q^2}\right)(\b q^2+\sigma_\chi^2)-2i\sigma_\chi\epsilon^{ijk}q_k\right]\,.
\gal
Its contribution to the cross section has the form $\mathcal{M}_i\mathcal{M}_j^*\aver{A^iA^{j*}}$. When summed over the particle helicities $\mathcal{M}_i\mathcal{M}_j^*$ is symmetric with respect to swapping the indices $i$ and $j$. This implies that the second term in the square brackets of \eq{Vc9} cancels out.\footnote{We note however, that this term  contributes to the partial cross sections for the helicity states.}

The photon propagator \eq{Wa5} has a resonance at $q^2=-\lambda \sigma_\chi |\b q|$ which in the static limit becomes $\b q^2= \sigma_\chi^2$ in \eq{Wb12c}. This resonant behavior can be regulated in the usual way by taking account of the finite resonance width:
\ball{Vf1}
\frac{1}{q^4+b^2q^2-(b\cdot q)^2}\to \frac{1}{q^4+b^2q^2-(b\cdot q)^2+i q^2\Gamma^2}\,,
\gal
so that the denominator now vanishes at $q^2=-\lambda \sigma_\chi |\b q|-i\Gamma^2/2$. 
It is noteworthy that the photon propagator exhibits both the $s$-channel ($q^2>0$) and the $t$-channel ($q^2<0$) resonances. This happens because $-\lambda \sigma_\chi |\b q|$ can be positive or negative depending on the photon polarization. The same parameter $\Gamma$ regulates both channels. We note  
that $\Gamma$ in the $s$-channel has a transparent physical meaning. Namely, it is related to the photon decay width $W$ as $\Gamma^2\approx \sigma_\chi  W$. The regularization procedure \eq{Vf1}  is slightly different than the one we used in \eq{Wj10}. In the denominator of the squared amplitude at $q^0=0$, the former produces $\b q^4+ \Gamma^4$, whereas the latter $\b q^2(\b q^2+E^2/(2E'^2\tau^2))$. 

The resonant behavior is closely related to the chiral magnetic instability of electromagnetic field in chiral medium which is driven by the modes with $\im q^0>0$ see e.g.\
\cite{Joyce:1997uy,Boyarsky:2011uy,Hirono:2015rla,Xia:2016any,Kaplan:2016drz,Kharzeev:2013ffa,Khaidukov:2013sja,Avdoshkin:2014gpa,Akamatsu:2013pjd,Kirilin:2013fqa,Tuchin:2014iua,Dvornikov:2014uza,Buividovich:2015jfa,Sigl:2015xva,Kirilin:2017tdh,Tuchin:2017vwb}. From the dispersion relation 
$(q^0)^2=\b q^2-\lambda \sigma_\chi  |\b q|$
it is evident that these modes have $\b q^2< \sigma_\chi ^2$. In the limit of small $q^0$ there is only one unstable mode $|\b q|=\sigma_\chi $. The instability is eventually tamed by the chirality flow between the magnetic field and the medium, which induces the time-dependence of $\sigma_\chi $. It is reasonable then to estimate $W$ as the inverse of the chirality transfer time $W\sim \alpha^2m^2/T$ \cite{Boyarsky:2011uy} ($T$ is the tempearture) which is the softest scale in the problem. 

The final result of the calculation in the ultrarelativistic limit is
\ball{Vj3}
\frac{d\sigma_\mu}{d\omega}\approx&\frac{ e^4 \textgoth{m} ^2  }{3(2\pi)^3 \omega }\frac{ E'}{E} \Bigg\{\left(\frac{(3\sigma_\chi ^2+2m^2)(E^2+E'^2)}{2m^2E'E}-\frac{E^2}{E^{'2}}-\frac{E'^2}{E^{2}}\right)\ln \frac{16 \epsilon^4 E'^4}{m^4 \omega ^4}\nonumber\\
&+4\left(\frac{E^2}{E'^2}  +\frac{E'^2}{E^2 }\right)-1
+\left(1-2\frac{E^2}{E'^2}\right) \ln \frac{4 E'^2}{m^2 }+\left(1-2\frac{E'^2}{E^{2}}\right) \ln \frac{4 E^{2}}{m^2 }
\nonumber\\
&+\left(\frac{E}{E'}+\frac{E'}{E}\right) \ln\frac{4E^2}{m^2}\ln\frac{4E'^2}{m^2}
  \nonumber\\
&+\frac{\sigma_\chi ^2m^2\pi}{\Gamma^2}\Bigg[\left(\frac{4}{E}+\frac{(E^2+E'^2)\sigma_\chi ^2}{m^2E^2E' }\right)\left(\frac{2E}{m^2}-\frac{\omega}{\sigma_\chi  E'}\right)-4\frac{E'^2}{E^3}  \left(\frac{E }{m^2}-\frac{\omega }{2\sigma_\chi  E' }\right)\nonumber\\
&-2\frac{E^2  }{ E'^3}\left(\frac{2E '}{m^2}-\frac{\omega }{\sigma_\chi  E }\right)+\frac{\omega^2}{m^2E^2}\left(\frac{\sigma_\chi (E^2+E'^2)}{E'^2\omega}-\frac{m^2E}{2E'^3}-\frac{m^2}{2E'E}\right)\Bigg] \nonumber\\
&\times\theta(\omega_0-\omega) \Bigg\}\,,
\gal
where the UV cutoff is 
\ball{Vj-3}
\omega_0=\frac{2E^2\sigma_\chi }{2E \sigma_\chi +m^2}\,.
\gal
In the limit $\sigma_\chi \to 0$ this equation reduces to the result  obtained by Gluckstern and Hull\cite{Gluckstern:1953zz}.

The two salient features of the anomalous contribution is the enhancement by the large factor $(b_0/\Gamma)^2$ as compared to the corresponding non-chiral expression \cite{Gluckstern:1953zz} and the emergence of the ultra-violet cutoff \eq{Vj-3} due to the resonance in the $t$-channel. However, the numerical calculation shown in \fig{Wfig:spectrum} and \fig{Wfig:spectrum2} indicates that  despite the resonance, the contribution of the magnetic dipole to the total bremsstrahlung cross section is small as compared to the Coulomb one due to smallness of the nuclear magnetic moment. 


\section{Summary}\label{sec:summary}

We discussed radiation and energy loss processes in the quark-gluon plasma in the framework of axion electro- and chromodynamics. The Chern-Simons term \eq{I1} stemming from the chiral anomaly\cite{Adler:1969gk,Bell:1969ts} takes account of the $CP$-odd axion excitations in chiral media. The main novel effect is the Chiral Cherenkov radiation whose magnitude competes with the conventional energy loss mechanisms. We also reviewed the effect of the axion excitations on the bremsstrahlung. The most distinctive features of all these effects is strong dependence on the gauge boson polarization. A comprehensive phenomenological analysis requires incorporation of the novel energy loss channels into the numerical packages describing jets in hot nuclear medium. 

The total photon spectrum emitted by the quark-gluon plasma is also sensitive to the chiral effects, especially in the infrared region. Many mechanisms that modify the photon spectrum in chiral media were discussed in the literature. Some of them also requite the presence of the external magnetic field \cite{Fukushima:2012fg,Wang:2024gnh}, but others do not \cite{Mamo:2015xkw,Mamo:2013jda,Tuchin:2019jxd}. 

While this review focuses mostly on spatially uniform axion excitations ($\b \nabla\theta=0$), the spatial inhomogeneities are very important in applications to Dirac and Weyl semimetals as they induce the anomalous Hall current. It was shown  that Chiral Cherenkov radiation exists in such systems as well \cite{Huang:2018hgk,Tuchin:2018mte} and in fact can be a source of technologically important polarized  THz radiation \cite{Hansen:2024kvc}. The spatial gradients are also important in the quark-gluon plasma if the size of the $CP$-odd domains is of the order a few fm or less. In this case, one has to impose appropriate boundary conditions on the fields at the domain walls. It was argued in \cite{Tuchin:2018rrw} that the boundary conditions strongly impact the dynamics of the chiral magnetic effect. Optical properties  of chiral domains in materials were discussed in \cite{Stewart:2019xjh} who derived the modified Fresnel equations in axion electrodynamics. 

Axion electrodynamics was used to investigate other chirality sensitive processes in the quark-gluon plasma. In \cite{Tuchin:2016tks} we considered radiation of the Chandrasekhar-Kendall states \cite{CK} by an ultrarelativistic quark scattering off  thermal gluons. These topological states are characterized by the magnetic helicity which can spontaneously change due to spatially inhomogeneities in the axion field \cite{Tuchin:2016qww}. 

This review by no means intends to review the entire field of axial electrodynamics, not even as it applies to the quark-gluon plasma. This subject is too large to fit in so few pages. Our hope is to convey to the reader some of the excitement that we have working on these subjects.


\section*{Acknowledgements}

This work was supported in part by the U.S. Department of Energy Grants No.\ DE-SC0023692.


\bibliographystyle{ws-ijmpe}

\bibliography{anom-biblio}

\end{document}